\documentclass[aps,prb,twocolumn,superscriptaddress,floatfix,letter]{revtex4-1}
\usepackage{graphicx}
\usepackage{physics}
\usepackage{amssymb}   
\usepackage{amsfonts}
\usepackage{amsmath}
\usepackage{dsfont}
\usepackage{dcolumn}   
\usepackage[makeroom]{cancel}
\usepackage[caption=false]{subfig}
\usepackage{subfig}
\usepackage{ulem}
\usepackage{bm}        
\usepackage[mathscr]{eucal}
\usepackage[dvipsnames]{xcolor}
\usepackage[colorlinks, linkcolor=OrangeRed,citecolor=RoyalBlue,urlcolor=NavyBlue]{hyperref}
\usepackage[all]{hypcap} 
\usepackage{mathtools}
\usepackage{wasysym}
\definecolor{mypurple}{rgb}{0.49,0.18,0.56}
\definecolor{mygold}{rgb}{0.93,0.49,0.13}

\newcommand{\canc}[1]{}
\begin{document}

\title{Local measures of dynamical quantum phase transitions}
\author{Jad C.~Halimeh}
\affiliation{INO-CNR BEC Center and Department of Physics, University of Trento, Via Sommarive 14, I-38123 Trento, Italy}

\author{Daniele Trapin}
\affiliation{Max-Planck-Institut f\"ur Physik komplexer Systeme, N\"othnitzer Stra{\ss}e 38,  01187-Dresden, Germany}

\author{Maarten Van Damme}
\affiliation{Department of Physics and Astronomy, University of Ghent, Krijgslaan 281, 9000 Gent, Belgium}

\author{Markus Heyl}
\affiliation{Max-Planck-Institut f\"ur Physik komplexer Systeme, N\"othnitzer Stra{\ss}e 38,  01187-Dresden, Germany}

\begin{abstract}
In recent years, dynamical quantum phase transitions (DQPTs) have emerged as a useful theoretical concept to characterize nonequilibrium states of quantum matter. DQPTs are marked by singular behavior in an \textit{effective free energy} $\lambda(t)$, which, however, is a global measure, making its experimental or theoretical detection challenging in general. We introduce two local measures for the detection of DQPTs with the advantage of requiring fewer resources than the full effective free energy. The first, called the \textit{real-local} effective free energy $\lambda_M(t)$, is defined in real space and is therefore suitable for systems where locally resolved measurements are directly accessible such as in quantum-simulator experiments involving Rydberg atoms or trapped ions. We test $\lambda_M(t)$ in Ising chains with nearest-neighbor and power-law interactions, and find that this measure allows extraction of the universal critical behavior of DQPTs. The second measure we introduce is the \textit{momentum-local} effective free energy $\lambda_k(t)$, which is targeted at systems where momentum-resolved quantities are more naturally accessible, such as through time-of-flight measurements in ultracold atoms. We benchmark $\lambda_k(t)$ for the Kitaev chain, a paradigmatic system for topological quantum matter, in the presence of weak interactions. Our introduced local measures for effective free energies can further facilitate the detection of DQPTs in modern quantum-simulator experiments.
\end{abstract}

\date{\today}
\maketitle
\tableofcontents
\section{Introduction} \label{introduction}
In the last decades the field of nonequilibrium quantum matter\cite{Taeuber_book,Eisert_review} has developed into a central research field in physics not only driven by fundamental theoretical questions but also by impressive experimental advances in various quantum-simulator platforms such as ultracold atoms\cite{Langen_review} or trapped ions.\cite{Leibfried_review} The level of control and precision available in modern experiments has facilitated the observation of various out-of-equilibrium phenomena such as many-body localization,~\cite{schreiber2015,smith2016,Choi2016} the quantum Kibble-Zurek mechanism,~\cite{2014Xu,2016Anquez,2016Clark,Cui2016,keesling2019quantum,2020Xue} gauge-theory dynamics,~\cite{Martinez2016,Goerg2019,Schweizer2019,Mil2019,Yang2020} many-body dephasing,\cite{Kaplan2020} and dynamical phase transitions.~\cite{jurcevic2016,Flaeschner2018,Zhang2017,Smale2019,Guo2019,Tian2020}

One key approach to characterize the resulting nonequilibrium quantum states has been to extend well-established concepts from equilibrium statistical physics\cite{Cardy_book,Sachdev_book} to the out-of-equilibrium regime, such as the notion of a local order parameter in the long-time steady state of a quantum many-body system, e.g, in the wake of a quench.~\cite{sciolla2011,Halimeh2017} Another extension has taken shape in the theory of dynamical quantum phase transitions~\cite{heyl2013,Heyl2017Review,Zvyagin2017Review} (DQPT). Whereas thermal phase transitions are connected to nonanalyticities at critical temperature in the thermal free energy,~\cite{huang2009, huang2016dynamical, lee1952statistical} 
a quantum many-body system during its temporal evolution can undergo a DQPT when the dynamical analog $\lambda(t)$ of the free energy exhibits nonanalyticities at \textit{critical evolution times} $t_c$. Much the same way as equilibrium phase transitions are controlled by a parameter such as temperature or pressure that can be properly tuned in an experiment, in the theory of DQPTs a quench parameter such as interaction or magnetic-field strength determines the presence of DQPTs or lack thereof. Moreover, evolution time $t$ can be understood as complexified inverse temperature in this analogy.\cite{heyl2013} A DQPT suggests that at the critical time, the state is drastically different from the initial one, and this difference is simply a consequence of the time evolution.
Furthermore, such a deviation from the initial state can be quantified with the associated information contained in the effective free energy $\lambda(t)$, also known as return probability or rate function. This object is a global quantity, which in general is difficult to access in experiments. In fact, such global measurements in quantum many-body systems require resources that scale exponentially in system size, leading naturally to the question of whether the essential information on DQPTs can also be obtained through less demanding measurements.

Guided by this experimental consideration, this work introduces two \textit{local} versions of the effective free energy: $\lambda_M(t)$ in real space and $\lambda_k(t)$ in momentum space, both of which can reliably and controllably detect DQPTs, with the added advantage that they can be experimentally obtained with much less effort compared to the effective free energy $\lambda(t)$. As we will show later, for specific quench protocols $\lambda(t)$ is a function of projectors over all lattice sites of the system, and consequently it is a global measure. With the aim of introducing the real-local effective free energy $\lambda_M(t)$, we consider projectors not on all the $N$ degrees of freedom of the chain but on only $M\ll N$ of them. Since DQPTs occur in the thermodynamic limit $N\to \infty$, the sharp feature emerging in $\lambda(t)$ at the critical time is smoothed out in $\lambda_M(t)$, but nevertheless one can still extract the critical behavior through a scaling analysis.~\cite{stanley1999scaling, nelson1975,pfeuty1970one} We test the validity and efficacy of $\lambda_M(t)$ in the nearest-neighbor Ising model, since the exact solution of the problem provides analytic results with which the numerical estimation of $\lambda_M(t)$ can be benchmarked.~\cite{Sachdev_book,  silva2008, huang2009,heyl2013, de2020entanglement}
As a further example we consider the Ising chain with power-law interactions in order to assess how well $\lambda_M(t)$ can predict the presence of an underlying DQPT in this context. Moreover, in such a setup we can focus on different kinds of DQPTs.~\cite{ ScalingHeyl, andraschko2014, 2017Karrasch, Trapin2018, homrighausen2017, zauner2017probing, halimeh2017dynamical, Halimeh2018a, halimeh2019dynamical} Interestingly, through scaling analysis of the real-local effective free energy $\lambda_M(t)$ at different configuration sizes $M$ we are able to extract universal critical exponents for the various DQPTs arising in the dynamics of the spin chains we consider. This is particularly useful in the quest for dynamical quantum universality classes, in which many open questions remain despite several recent studies within the framework of DQPT.\cite{gurarie2019dynamical,halimeh2019dynamical,wu2019dynamical,wu2020nonequilibrium,wu2020dynamical}

In case experimental measurements involve operators defined in momentum space, it is more suitable to use the momentum-local effective free energy $\lambda_k(t)$ to detect DQPTs. We demonstrate this measure in an \textit{interacting Kitaev chain}\cite{Katsura2015} (IKC) representing a paradigmatic model for topological quantum matter. As a first step, we motivate the introduction of $\lambda_k(t)$, which conveniently relies on the fact that the effective free energy in the noninteracting Kitaev chain\cite{Kitaev2001} can be derived exactly in terms of local observables in $k$-space.

The theory of DQPT was developed to establish a mathematical background to concepts such as phase transitions in the dynamical regime, where tools and quantities provided by statistical physics are not applicable.
This necessitates introducing new objects, such as the Loschmidt amplitude~\cite{heyl2013, obuchi2017complex}
\begin{equation}
    \mathcal{L}(t) = \bra{\psi} e^{-iHt}\ket{\psi},
\end{equation}
which quantifies the overlap of the time-evolved state $\ket{\psi(t)}= e^{-iHt}\ket{\psi}$ from the initial one $\ket{\psi }$.
The structure of the Loschmidt amplitude resembles the boundary partition function defined in statistical mechanics, but the time-evolution operator makes it a complex quantity instead of real. This analogy suggests the introduction of the effective free energy 
\begin{equation}\label{lambda}
    \lambda(t) = -\lim_{N\to\infty}\frac{1}{N} \ln\lvert\mathcal{L}(t)\rvert^2,
\end{equation}
where $N$ is the number of degrees of freedom.
Motivated by equilibrium physics where phase transitions are defined at those values of the control parameter that make the free energy nonanalytic, similarly DQPTs occur at the time $t_c$ when the effective free energy shows a nonanalytic cusp.
In order to observe a DQPT, a specific protocol must be adopted to bring the system out of equilibrium.~\cite{sharma2015quenches, sharma2016slow} Several ways have been studied but the most common one consists of a global quantum quench, where an out-of-equilibrium initial state is suddenly quenched by a given Hamiltonian. Usually, the initial state is prepared as the ground state of an initial Hamiltonian, a control parameter of which is then subsequently suddenly switched to a different value thereby realizing the final Hamiltonian.\cite{Mori_review}

The equilibrium and dynamical realms do not have in close analogy only the definition of phase transitions, but other important properties have been found in both regimes.
For example, through RG analysis it has been found that the DQPTs emerging from quenching the system with the classical nearest-neighbor Ising chain show scaling and universality, clear features of continuous phase transitions.~\cite{heyl2013, ScalingHeyl, Trapin2018} When the range of interactions is extended to realize a long-range quantum Ising chain, differences between the equilibrium and dynamical regimes arise. In particular, the dynamical phase diagram fundamentally differs from the equilibrium one,\cite{Halimeh2018a} and in particular, a new type of cusp in the effective free energy appears that does not correspond to any zeros in the dynamics of the order parameter, thereby leading to the definition of \textit{anomalous} DQPTs.~\cite{halimeh2017dynamical, defenu2019dynamical, zauner2017probing, homrighausen2017} Although these anomalous cusps have not been investigated in experiments yet, a lot of experimental achievements have been reached so far in the field. Using Rydberg atoms~\cite{bernien2017probing,browaeys2020many,labuhn2016tunable, kim2018detailed, barredo2015coherent,2017Marcuzzi} it is in fact possible to mimic the time evolution of spin chains under Ising Hamiltonians whose interaction range can be tuned properly. Employing instead ultracold-atom platforms,~\cite{Flaeschner2018} DQPTs in topological systems have been explored, quenching the system from different topological classes.~\cite{budich2016dynamical,HeylBudich2017,bhattacharya2017mixed,huang2016dynamical,qiu2018dynamical,Haifeng2018,hagymasi2019dynamical} In this context, it is possible to relate the DQPT with the so-called topological dynamical order parameter,\cite{budich2016dynamical,Zache2019} extending the bridge between equilibrium and out-of-equilibrium, where the introduction of an order parameter on general grounds is currently a major challenge.

The manuscript is organized as follows. In Sec.~\ref{model} we motivate the reasons why we are interested in introducing new \textit{local} quantities to detect DQPTs. In particular, when experimental measurements concern spin degrees of freedom, it is more natural to define such a quantity in real space, called real-local effective free energy $\lambda_M(t)$. On the other hand, when observables are defined in momentum space, it is sensical to use a tool which is also defined in this framework. Such a quantity is named momentum-local effective free energy $\lambda_k(t)$.
Subsequently, we describe the models used to test both $\lambda_M(t)$ and $\lambda_k(t)$. In the former case we consider the nearest-neighbor and long-range quantum Ising chains. For the latter case, we choose the Kitaev chain with small interactions. In Sec.~\ref{RL_lambda}, we mathematically rigorously introduce the real-local effective free energy $\lambda_M(t)$, and provide results showing its efficacy in detecting DQPTs in suitable models. A similar analysis is carried out in Sec.~\ref{ML_lambda}, where we focus instead on the momentum-local effective free energy $\lambda_k(t)$. We summarize our findings in Sec.~\ref{conclusions}, and furthermore we provide supplemental results in Appendix~\ref{sec:extra}, specifications of our numerical implementations in Appendix~\ref{numspecs}, and derivational details in Appendix~\ref{mapping}.

\section{Experimentally accessible quantities to detect DQPTs} \label{model}
The main goal of this manuscript is the introduction of two quantities for the reliable detection of DQPTs that require a significantly reduced amount of resources as compared to the Loschmidt amplitude. Depending on the particular system under consideration, it might be more convenient to consider quantities in either real space or in momentum space. As a consequence, we consider both scenarios in illustrating the efficacy of the real-local effective free energy $\lambda_M(t)$ and the momentum-local effective free energy $\lambda_k(t)$. The full details of their formal definition will be provided respectively in Secs.~\ref{RL_lambda} and~\ref{ML_lambda}, while for the moment we give a brief overview of both measures. As opposed to the Loschmidt echo $\lvert\mathcal{L}(t)\rvert^2$ which is a global quantity, the real-local Loschmidt echo $\lvert\mathcal{L}_M(t)\rvert^2$ is not since it is obtained as the expectation value of a projector $\hat{P}_M$ onto a given \textit{local} real-space configuration at $M$ (usually adjacent) sites on the lattice, where $M$ is finite and therefore considered much smaller than the system size $N$. The aforementioned local configuration can be conveniently chosen on the associated $M$ sites as the product state that is closest to the initial state of the quench protocol under consideration. On general grounds, we can therefore write $\lvert\mathcal{L}_M(t)\rvert^2 = \bra{\psi(t)}\hat{P}_M\ket{\psi(t)}$, where $\ket{\psi(t)}$ is the time-evolved state. In particular, when $M=N$ and the initial state is the configuration onto which $\hat{P}_M$ projects, $\lambda_M(t)=\lambda(t)$ becomes exact. In practice, the useful range of values that $M$ can assume is given by the tradeoff between being small enough in order to feasibly measure the real-local Loschmidt echo $\lvert\mathcal{L}_M(t)\rvert^2$ in an experiment, and large enough to observe a clear signature of the underlying DQPT in the real-local effective free energy $\lambda_M(t)$. As we will show in Sec.~\ref{RL_lambda}, scaling behavior allows us to surmise the presence of a DQPT from a few small values of $M$. The real-local effective free energy $\lambda_M(t)$ is useful in the context of spin Hamiltonians, such as nearest-neighbor or power-law interacting Ising chains, where typical experimental measurements involve spin degrees of freedom on a lattice in real space.

On the other hand, the momentum-local effective free energy $\lambda_k(t)$ is given by the product of momentum uncorrelated two-body correlation functions. In general, the effective free energy $\lambda(t)$ can be formulated in terms of  $n$-point functions in momentum space. Our construction of $\lambda_k(t)$ captures the information of the underlying DQPT contained only in two-point functions, which are easily accessible in modern ultracold-atom and ion-trap experiments.\cite{Flaeschner2018} In case of, e.g., a two-band free fermionic model, $\lambda_k(t)=\lambda(t)$ is exact. In more generic models where experimental measurements concern mainly observables defined in momentum space, it is natural to study the behavior of the momentum-local effective free energy $\lambda_k(t)$ to investigate the emergence of DQPTs as an adequate approximation to $\lambda(t)$. This situation arises, for example, in topological systems where one typically measures expectation values of fermionic operators defined in momentum space.

\section{Real-local effective free energy $\lambda_M(t)$} \label{RL_lambda}
Because the Hilbert space of a generic quantum many-body model scales exponentially in system size, measurement of the effective free energy $\lambda(t)$ due to dynamics actuated by such a model will also require a number of resources exponentially large in system size. We are therefore interested in defining quantities that can be efficiently obtained with significantly fewer resources, yet that are still able to reliably detect DQPTs.

\subsection{Nearest-neighbor classical Ising model}\label{NN}
We take a first step in this direction by introducing the real-local effective free energy $\lambda_M(t)$ and testing it for a particular quench protocol.
We consider as initial state a chain where all spins point along the positive $z$-direction: $\ket{\psi} = \ket{\uparrow_1 ... \uparrow_N}$, which is a paramagnetic product state.
When the quench is performed, the system undergoes time evolution propagated by the classical nearest-neighbor Ising model given by the Hamiltonian

\begin{align}
H=-J\sum_{j=1}^N\sigma_j^x\sigma_{j+1}^x,
\end{align}
where $\sigma_j^a,\,a=x,y,z$ are the Pauli matrices on site $j$, $J=1$ sets the energy scale, and $N$ is the number of sites. First we examine the Loschmidt echo $\lvert\mathcal{L}(t)\rvert^2$, which is a global measure since it can be written in terms of projectors over all spins of the system:
\begin{align}\nonumber
		\lvert\mathcal{L}(t)\rvert^2 &= \lvert\bra{\psi} e^{-iHt} \ket{\psi}\rvert^2 =\bra{\psi(t)}\ket{\psi}\bra{\psi}\ket{\psi(t)}\\\label{LE}
		&=\bra{\psi(t)}\prod_{j=1}^N \hat{p}_j^z \ket{\psi(t)},
\end{align}
where $\hat{p}_j^z=\ket{\uparrow_j }\bra{\uparrow_j}$ is the local projector onto the state $\ket{\uparrow_j}$ on site $j$. In the case of a fully $z$-polarized initial product state as we consider here, the Loschmidt echo reduces to
\begin{align}\label{LE_NN}
\lvert\mathcal{L}(t)\rvert^2 =  \left\lvert(\cos t)^N + (\sin t)^N \right\rvert^2,
\end{align}
leading to the effective free energy
\begin{align}\nonumber
\lambda(t)&= -\lim_{N\to\infty}\frac{2}{N} \ln \left\lvert(\cos t)^N + (\sin t)^N \right\rvert\\
&=\begin{cases}
-2\ln\lvert\cos t\rvert, & \text{if}\,\,\,\lvert\cos t\rvert\geq\lvert\sin t\rvert, \\
-2\ln\lvert\sin t\rvert, & \text{if}\,\,\,\lvert\cos t\rvert<\lvert\sin t\rvert.
\end{cases}
\end{align}
Therefore the DQPT occurs at $t_c=\pi/4$ when the two terms in Eq.~\eqref{LE_NN} are equal.

In order to construct a local version of the Loschmidt echo, we define $\hat{P}_M^z$ as a projector on a finite \textit{configuration} of $M$ lattice sites: 
\begin{equation}
		\hat{P}_M^z :=	\prod_{j=1}^M \hat{p}_j^z.
\end{equation}

\begin{figure}[ht!]
	\centering
	\includegraphics[width=1.0\columnwidth]{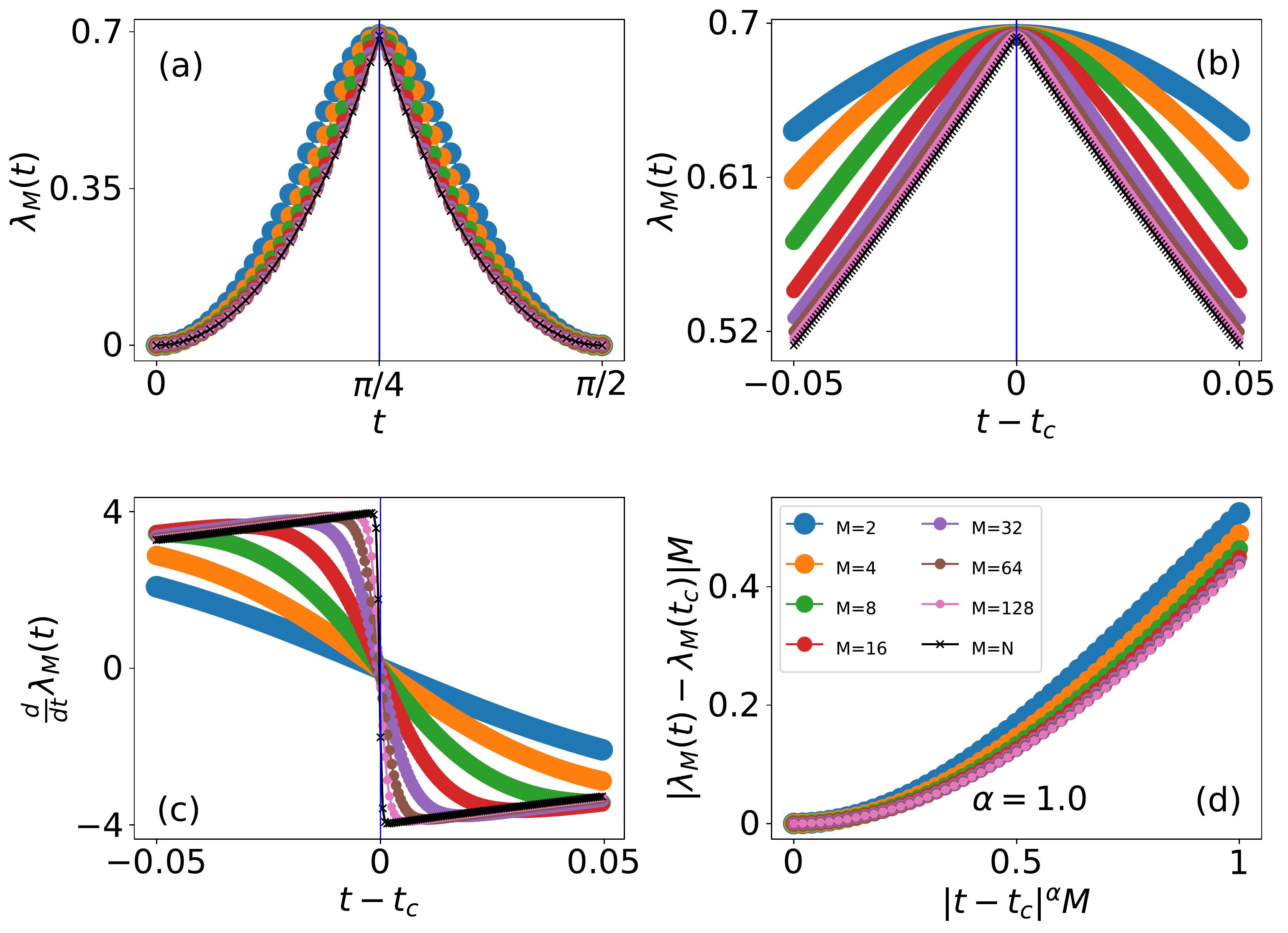}
	\caption{(a) Effective free energy $\lambda(t)$ (black crossed line) and real-local effective free energy $\lambda_M(t)$ (colored dotted lines) as a function of time $t$ for different values of $M=2, \;4,\; 8,\; 16,\; 32,\; 64,\; 128$ sites. Recall that $\lambda_{M=N}(t)=\lambda(t)$. We find that even when $M\ll N\to\infty$, $\lambda_M(t)$ reliably detects the DQPT, even yielding a practically exact estimate of the critical time $t_c=\pi/4$. (b) Zoom-in of $\lambda_M(t)$ as a function of $t-t_c$ around the critical time. (c) Time-derivative of real-local free energy, $d\lambda_M(t)/dt$, as a function of $t-t_c$. As $M$ increases, the discontinuity  jump at the critical time $t_c$ becomes sharper. (d) $\lvert\lambda_M(t) - \lambda_M(t_c)\rvert M$ as a function of $\lvert t-t_c\rvert M$. The plot shows good collapse of the curves for different $M$, particularly when $t$ is close to $t_c$, and reveals a critical exponent $\alpha=1$.}
	\label{fig:local}
\end{figure}

We are now in the position to define the real-local Loschmidt echo as 
\begin{align}
\lvert\mathcal{L}_M(t)\rvert^2 = \bra{\psi(t)}\hat{P}_M^z\ket{\psi(t)},
\end{align}
and consequently the associated real-local effective free energy
\begin{align}
\lambda_M(t)= -\frac{1}{M}\ln\lvert\mathcal{L}_M(t)\rvert^2.
\end{align}
We note that in the limit of $M=N$, we restore the full Loschmidt echo $\lvert\mathcal{L}_{M=N}(t)\rvert^2 = \lvert\mathcal{L}(t)\rvert^2$, and thus the associated real-local effective free energy $\lambda_{M=N}(t)=\lambda(t)$ is exact.

We show in Fig.~\ref{fig:local}(a) the real-local return probability $\lambda_M(t)$ for different values of $M=2,\,4,\,8,\,16,\,32,\,64,\,128$ sites, along with the exact effective free energy $\lambda(t)$. In Fig.~\ref{fig:local}(b) we provide a zoom-in around the cusp occurring at the critical time $t_c = \pi/4$. The sharp feature is visible for large $M$, while $\lambda_M(t)$ becomes smoother with decreasing $M$. Interestingly, $\lambda_M(t)$ shows a maximum at $t=t_c$, indicating accurate estimation of the critical time of the DQPT. In order to better outline such behavior, we show in Fig.~\ref{fig:local}(c) the time derivative of the real-local effective free energy $d\lambda_M(t)/dt$.
In this case, the nonanalyticity occurring at the critical time is more evident in the limit of large $M$ and manifests itself as a discontinuity in $d\lambda_M(t)/dt$. Much the same way as for generic phase transitions in equilibrium, DQPTs also occur generically only in the thermodynamic limit (see, e.g., Ref.~\onlinecite{Puebla2020} for a specific counterexample), i.e., $N \rightarrow \infty$, and a finite local configuration of $M$ sites will in general not allow $\lambda_M(t)$ to exhibit a nonanalyticity. Nevertheless, finite-size scaling analysis overcomes the problem of detecting phase transitions when dealing with finite $M$.

Renormalization group\cite{Wilson1971a,Wilson1971b} (RG) analysis shows that for the quench considered the DQPT occurring at critical time $t_c$ is continuous.~\cite{ScalingHeyl, Trapin2018}
 As a consequence, we expect scaling and universality, such that the singular part of the return probability in the vicinity of the critical time assumes the form
\begin{equation}
	\lambda(t\sim t_c) \sim N^{-\alpha/\nu} f_{\lambda}(N \Delta t^{\nu}),
	\label{lambda_scaling}
\end{equation}
which is inspired from scaling behavior in equilibrium,\cite{Sachdev_book} with $\Delta t = \lvert t-t_c\rvert$, $f_{\lambda}$ a universal scaling function, $\alpha$ is an \textit{a priori} generic critical exponent, while $\nu$ is the one related to the free energy and also the critical exponent of the correlation length in thermal equilibrium,\cite{Cardy_book} $\xi \sim \lvert T/T_c-1\rvert^{-\nu}$, where $T$ is temperature and $T_c$ is its critical value. In our case we expect $\alpha=\nu$, since both the free energy at equilibrium and the return probability in the DQPT theory are proportional to the logarithm of the same partition function, which is only real in the former case, while complex in the latter.

In the quench considered here, RG analysis indicates $\nu=1$.~\cite{ScalingHeyl, Trapin2018}
Taking all these considerations into account, we expect therefore that
\begin{equation}
	\lambda(t\sim t_c) \sim \frac{1}{N} f_{\lambda}(N \Delta t).
	\label{lambda_3}
\end{equation}
In general, one can use Eq.~\eqref{lambda_scaling} when the cutoff scale is $N$ and not $\Delta t^{\nu}$, meaning that $ N< \Delta t^{-\nu}$.
After defining $x=N\Delta t$, the condition required to use Eq.~\eqref{lambda_scaling} for our case reads 
\begin{equation} \label{range_validity}
	N < \frac{1}{\Delta t} \rightarrow	x < 1.
\end{equation}
The scaling function of Eq.~\eqref{lambda_3} suggests that by plotting  $\lvert\lambda - \lambda(t_c)\rvert N$ as a function of $\lvert t-t_c\rvert N$, the curves for different system sizes $N$ collapse onto each other.

When we consider the real-local return probability $\lambda_M(t)$,
the parameter $M$ represents the inverse distance to the critical point, which is at $M\rightarrow\infty$. As a consequence, we have to update the scaling function in order to account for this, leading to
\begin{align}\nonumber
	\lambda_M(t) &\sim \frac{1}{M}f_{\lambda_M}\left( \frac{M}{N}, M \Delta t \right)\\
	&\underset{M\ll N}{\overset{}{\longrightarrow}} 
	\frac{1}{M}g_{\lambda_M}\left(M \Delta t \right),
	\label{lambda_M}
\end{align}
where $f_{\lambda_M}$ and $g_{\lambda_M}$ are universal scaling functions. One can show that for the quench considered here, in the range $M<2N$, $\lambda_M(t)$ assumes the form
\begin{equation}
	\lambda_M(t) = -\frac{1}{M} \ln \left\lvert(\cos t)^{2(M+1)} + (\sin t)^{2(M+1)} \right\rvert.
\end{equation}
We compute a Taylor expansion of $\lambda_M(t)$ close to the critical time to obtain
\begin{equation}
	\lambda_M(t\sim t_c) \simeq \ln2 - 2(M+1)(t-t_c)^2.
	\label{lambda_2}
\end{equation}
From the Taylor expansion in Eq.~\eqref{lambda_2} we notice that $\lambda_M(t=t_c) = \ln2$. Furthermore, in the limit of $M  \gg 1$ Eq.~\eqref{lambda_2} can be rewritten as
\begin{equation}
	\lvert\lambda_M(t) - \lambda_M(t_c)\rvert M \simeq 2\big(M \lvert t-t_c\rvert\big)^2,
	\label{lambda_4}
\end{equation}
which is consistent with the scaling prediction outlined in Eq.~\eqref{lambda_M}. In Fig.~\ref{fig:local}(d) we plot $\lambda_M(t)$ according to the prescription in Eq.~\eqref{lambda_M}: $\lvert\lambda_M(t) - \lambda_M(t_c)\rvert M$ as a function of $\lvert t-t_c\rvert M$. 
In the small $x=\lvert t-t_c\rvert M$ limit, the curves for different values of $M$ collapse onto each other exhibiting a parabolic behavior as suggested by the Taylor expansion in Eq.~\eqref{lambda_4}. Towards the end of validity of the scaling function~\eqref{lambda_M}, $x\rightarrow 1$, some deviations between the curves appear, in particular when $M$ is small.

\subsection{Long-range transverse-field Ising chain}\label{long}
We now further probe the efficacy of the real-local effective free energy $\lambda_M(t)$ by considering long-range quantum Ising chains. Our theoretical interest is based on the experimental fact that Rydberg atom platforms~\cite{bernien2017probing,browaeys2020many,labuhn2016tunable, kim2018detailed, barredo2015coherent,2017Marcuzzi} and systems of trapped ions~\cite{jurcevic2016} can realize quench dynamics in the long-range transverse-field Ising model given by the Hamiltonian
\begin{align}\label{H_long}
	H = - \sum_{m<n} J_{m,n} \sigma_m^x \sigma_n^x - h \sum_{m=1}^N \sigma_m^z.
\end{align}
In experiments, the spin-spin coupling can be tuned to be of the kind $J_{m,n}\sim \lvert m-n\rvert^{-\mu}$ in the limit of large distance $\lvert m-n\rvert \gg 1$.
While for Rydberg atom architectures typical exponents are either $\mu=3$ and $\mu=6$,\cite{bernien2017probing,browaeys2020many} in systems of trapped ions $\mu$ can range from $0$ to $3$.\cite{Lanyon2011,Islam2013,Jurcevic2014,jurcevic2016,Neyenhuis2017} For the sake of numerical feasibility, we will assume that $J_{m,n}=J\lvert m-n\rvert^{-\mu}$ for any distance $\lvert m-n\rvert\geq1$ rather than only when $\lvert m-n\rvert\gg1$, with $J=1$ setting the energy scale. This is nevertheless a good approximation for the aforementioned experimental implementations, especially when the dynamical properties unique to long-range interactions are due to the interaction-profile tails.\cite{Halimeh2017,Liu2019} 
	
In equilibrium, this model exhibits a rich phase diagram, with the equilibrium quantum critical point $h_c^e(\mu)$ being $\mu$-dependent. Whereas for $\mu\geq3$ it falls in the short-range universality class, for $\mu<5/3$ mean-field analysis is exact.\cite{Knap2013} For $\mu<2$, the power-law interacting quantum Ising chain hosts a finite-temperature phase transition,\cite{Landau2013,Dyson1969,Thouless1969,Dutta2001} while at $\mu=2$ the model exhibits a Berezinskii-Kosterlitz-Thouless (BKT) transition.\cite{Dutta2001} 

The dynamical phase diagram of the long-range quantum Ising chain is also quite rich. A dynamical critical point\cite{vajna2014,andraschko2014,Jafari2019} $h_c^d(\mu,h_i)$ emerges that is dependent on both $\mu$ and the initial condition $h_i$ at which the quench starts, where for $h_i\leq h_c^e(\mu)$ numerical studies\cite{halimeh2017dynamical,zauner2017probing,homrighausen2017,lang2018} suggest that $h_c^d(\mu,h_i)<h_c^e(\mu)$, while for $h_i>h_c^e(\mu)$, the dynamical and equilibrium critical points are the same, $h_c^d(\mu,h_i)=h_c^e(\mu)$. Generically, the dynamical critical point $h_c^d(\mu,h_i)$ is the value of transverse-field strength across which the quench from $h_i$ must be carried out in order to see \textit{regular} DQPTs in the effective free energy. At the same time, particularly in the presence of sufficiently long-range interactions and for $h_i<h_c^e(\mu)$, $h_c^d(\mu,h_i)$ separates between an ordered long-time steady state for $h_f<h_c^d(\mu,h_i)$, and a paramagnetic long-time steady state for $h_f>h_c^d(\mu,h_i)$.\cite{zunkovic2016,halimeh2017dynamical,homrighausen2017,lang2018} Indeed, in the fully connected limit $\mu=0$ of Hamiltonian~\eqref{H_long}, a closed-form expression can be found for the dynamical critical point, where starting in the ordered phase $h_i<h_c^e(\mu=0)$ at zero preparation temperature for example, $h_c^d(\mu=0,h_i)=[h_c^e(\mu=0)+h_i]/2$ upon Kac-normalizing the interaction term.\cite{Kac1963} This dynamical critical point then separates a ferromagnetic long-time steady state and an effective free energy dominated by anomalous DQPTs for $h_f<h_c^d$, from a paramagnetic long-time steady state and an effective free energy dominated by regular DQPTs for $h_f>h_c^d$. This has been shown in exact numerics\cite{homrighausen2017,lang2018} and through an analytic semiclassical approximation.\cite{lang2018dynamical}

One of the intriguing out-of-equilibrium phenomena of the model in Eq.~\eqref{H_long} is that it allows for the effective free energy $\lambda(t)$ to exhibit different kinds of cusps, depending on the quench protocol employed. Generically, quenches from the ordered phase, $h_i<h_c^e(\mu)$, across the dynamical critical point $h_c^d(\mu,h_i)$ give rise to \textit{regular} DQPTs that are connected to zeros in the order-parameter dynamics.\cite{heyl2013,zunkovic2016,halimeh2017dynamical} In systems with finite-range interactions, such as nearest- or next-nearest-neighbor interactions, quenches within the ordered phase generically give rise to a fully analytic effective free energy. However, when the interactions are \textit{expansive}, such as, e.g., power-law or even exponentially decaying, if the quench Hamiltonian contains local ``spin-flip'' excitations as its lowest-lying quasiparticles,\cite{Liu2019,defenu2019dynamical} \textit{anomalous} DQPTs can emerge in the effective free energy even when the order paremeter does not change sign during the time evolution. Such anomalous signatures have been seen in one-dimensional power-law\cite{halimeh2017dynamical,zauner2017probing} and exponentially\cite{Halimeh2018a} decaying transverse-field Ising chains, two-dimensional quantum Ising models,\cite{Hashizume2018,Hashizume2019} and the Lipkin-Meshkov-Glick (LMG) model at zero\cite{homrighausen2017} and finite\cite{lang2018} temperature. In the following, we will consider these different kinds of DQPTs, and analyze the reliability of the real-local effective free energy in discerning them. For numerical feasibility we shall consider in our analysis only the first or second DQPT that arises in the effective free energy.

Our numerical results are calculated using infinite matrix product states\cite{Fannes1992,Verstraete2008,Zauner2018,MPSKit} (iMPS) based on the time-dependent variational principle\cite{Haegeman2011,Haegeman2013,Haegeman2016} (TDVP). In particular, we benchmark our results using two independent implementations of the same method. Both toolkits give identical results within machine precision. We achieve convergence at a maximal bond dimension $D=400$--$450$ for the regular cusps we consider in this work, and $D=450$--$500$ for their anomalous counterpart. We add supplemental results to this model in Appendix~\ref{sec:extra}, and discuss the implementation of Hamiltonian~\eqref{H_long} with power-law interactions in Appendix~\ref{numspecs}.

\subsubsection{Quenches from the paramagnetic phase}

\begin{figure}[ht!]
	\centering
	\includegraphics[width=1.0\columnwidth]{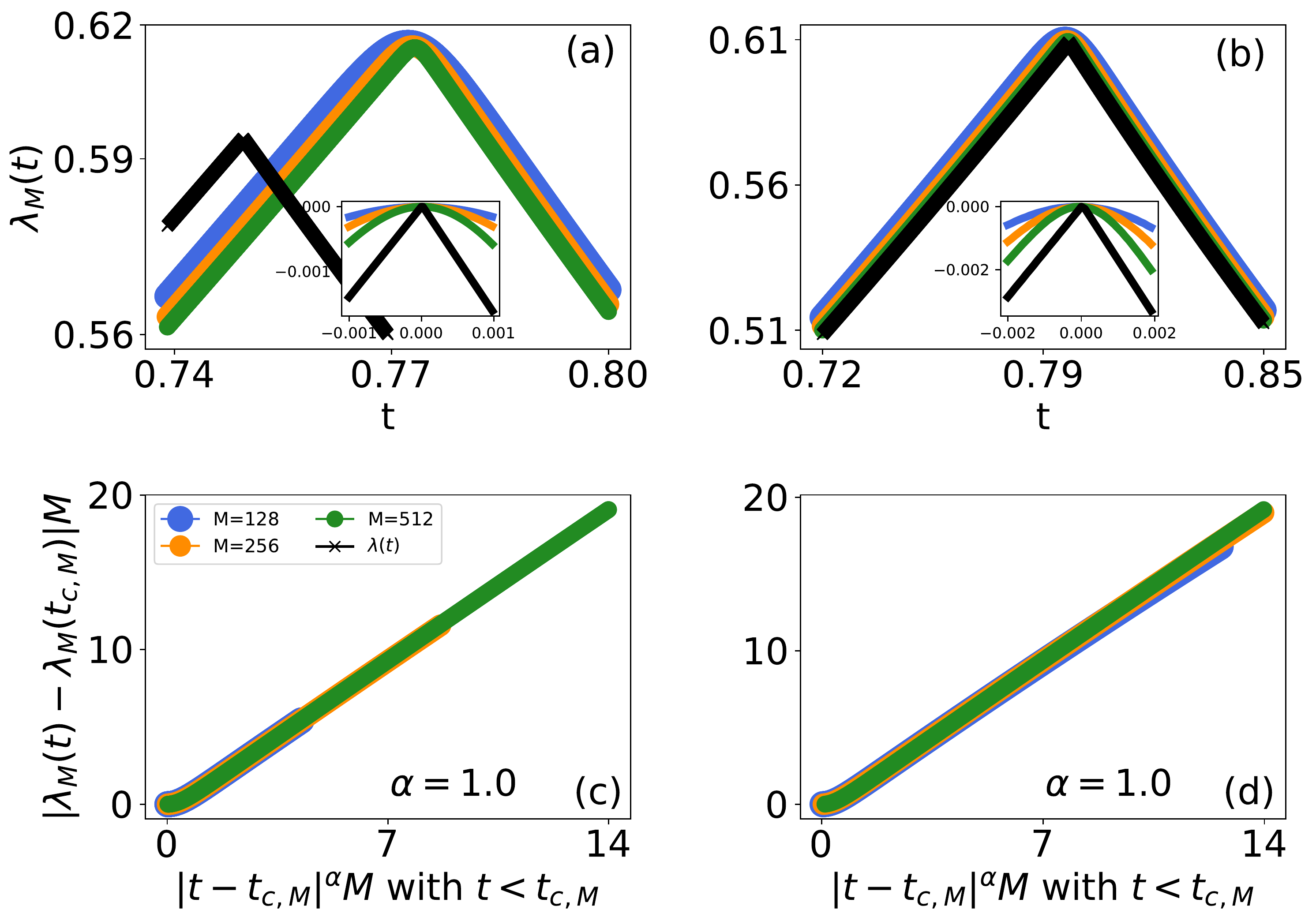}
	\caption{DQPTs arising in the long-range quantum Ising chain given by the Hamiltonian~\eqref{H_long} at $\mu=2$ in the wake of a quench from the paramagnetic [$h_i>h_c^e(\mu=2)$] to the ferromagnetic phase [$h_f=0.25<h_c^e(\mu=2)\approx2.5$]. The initial value of the transverse-field strength is $h_i=5$ in (a,c) and $h_i\to\infty$ for (b,d). In (a,b) we show the real-local effective free energy $\lambda_M(t)$ as a function of evolution time $t$ (colored dotted lines), while the black crossed line represents the effective free energy $\lambda(t)$. The insets contain a zoom-in of the sharp feature occurring at the critical time, showing $\lambda_M(t)-\lambda_M(t_{c,M})$ as a function of $t-t_{c,M}$, where $t_{c,M}$ is the estimated critical time from $\lambda_M(t)$, taken as the time of the corresponding peak in the latter. Note how the estimate $t_{c,M}$ is off in the case of $h_i=5$, while it is very accurate in the case of $h_i\to\infty$, and this is because in the latter case the configuration projected on in $\lambda_M(t)$ has the same spin alignment as the ground state of Hamiltonian~\eqref{H_long} at $h_i\to\infty$. In (c,d) we present the scaling analysis of $\lambda_M(t)$ shown in (a,b), respectively. The collapse is obtained by plotting $\lvert\lambda_M(t) - \lambda_M(t_{c,M})\rvert M$ as a function of $\lvert t-t_{c,M}\rvert M$. Despite having the same critical exponent $\alpha=1$, the scaling function here is linear, unlike the case of the classical Ising model where it is quadratic; cf.~Fig.~\ref{fig:local}.}
	\label{fig:regular}
\end{figure}

The dynamical phase diagram of the power-law quantum Ising chain shows that DQPTs appear in the effective free energy $\lambda(t)$ for quenches from the paramagnetic phase to the ferromagnetic phase,~\cite{halimeh2017dynamical} just as in the case of the nearest-neighbor quantum Ising chain.\cite{ScalingHeyl} Accordingly, we perform two quenches starting in the paramagnetic ground state of Eq.~\eqref{H_long} with interaction exponent $\mu=2$ at initial transverse field-strength values $h_i=5\approx2h_c^e(\mu=2)$ and $h_i\to\infty$, and ending in its ferromagnetic phase at final value $h_f=0.25<h_c^e(\mu=2)$ of the transverse-field strength.

The ensuing dynamics of the effective free energy $\lambda(t)$ and its approximation, the real-local effective free energy $\lambda_M(t)$, are shown in Fig.~\ref{fig:regular}(a,b) for the quench starting at $h_i=5$ and $h_i\to\infty$, respectively. Both panels show the effective free energy (black crossed line) exhibiting a DQPT, with critical time $t_c\approx0.75$ for $h_i=5$ in panel (a) and $t_c\approx0.79$ for $h_i\to\infty$ in panel (b). For the range of configuration sizes, $M=128,\, 256,\,512 $ sites, that we use for the real-local effective free energy $\lambda_M(t)$, the case of $h_i\to\infty$ shows a sharper feature at $t\approx0.79$ with larger $M$ (see insets). The approximate critical time $t_{c,M}$ predicted by $\lambda_M(t)$ is taken as the time of the corresponding peak in $\lambda_M(t)$, which sharpens into an actual DQPT at $M\to\infty$. As we see in Fig.~\ref{fig:regular}(a), $t_{c,M}$ nontrivially differs from the actual critical time $t_c$ obtained from $\lambda(t)$, and does not approach it with increasing $M$. This is due to fact that the initial state for the quench starting at $h_i=5$ does not consist of a chain with all spins aligned along the $z$-direction, which is the alignment configuration used in the projection employed in Eq.~\eqref{LE} for quenches starting in the paramagnetic phase. Indeed, in panel (b) where the initial field is fully $z$-polarized, the critical times $t_{c,M}$ and $t_c$ predicted by $\lambda_M(t)$ and $\lambda(t)$, respectively, are almost the same, with $t_{c,M}\to t_c$ with increasing $M$. In principle, the projection in Eq.~\eqref{LE} can be generalized to a configuration that can accommodate any initial state, such as the partial trace of the ground state of Eq.~\eqref{H_long} at $h_i=5$ along $M$ sites, and this would allow for a better estimation of the critical point through $\lambda_M(t)$. However, as Fig.~\ref{fig:regular}(a) shows, this is not necessary to detect a signature of DQPT, albeit it may be crucial when an accurate estimation of the critical time is desired.

We perform a scaling analysis for the results in Fig.~\ref{fig:regular}(a,b) and show the corresponding collapse at times around the DQPT in Fig.~\ref{fig:regular}(c,d), respectively. This is done upon rescaling the $y$-axis as $\lvert \lambda_M(t) - \lambda_M(t_{c,M})\rvert M$ and the $x$-axis as $\lvert t-t_{c,M}\rvert^\alpha M$. We show only scaling-analysis results for times $t<t_{c,M}$ (as we do throughout the whole paper) since iMPS results at earlier times are numerically always more accurate, and because we have also checked that the same universal scaling behavior occurs at $t>t_{c,M}$ within the precision of our results. We obtain the best collapse upon setting the critical exponent $\alpha=1$. This is the same result observed with the nearest-neighbor Ising model described in Sec.~\eqref{RL_lambda}. We have checked that our conclusions are not restricted to the case of $h_f=0.25$, and hold for other values of $h_f<h_c^e(\mu=2)$. In particular, we have obtained the same critical exponent when repeating the above quenches for $h_f=0$ (not shown).

\subsubsection{Quenches from the ordered phase: regular DQPTs}\label{regular}

\begin{figure}[ht!]
	\centering
	\includegraphics[width=1.0\columnwidth]{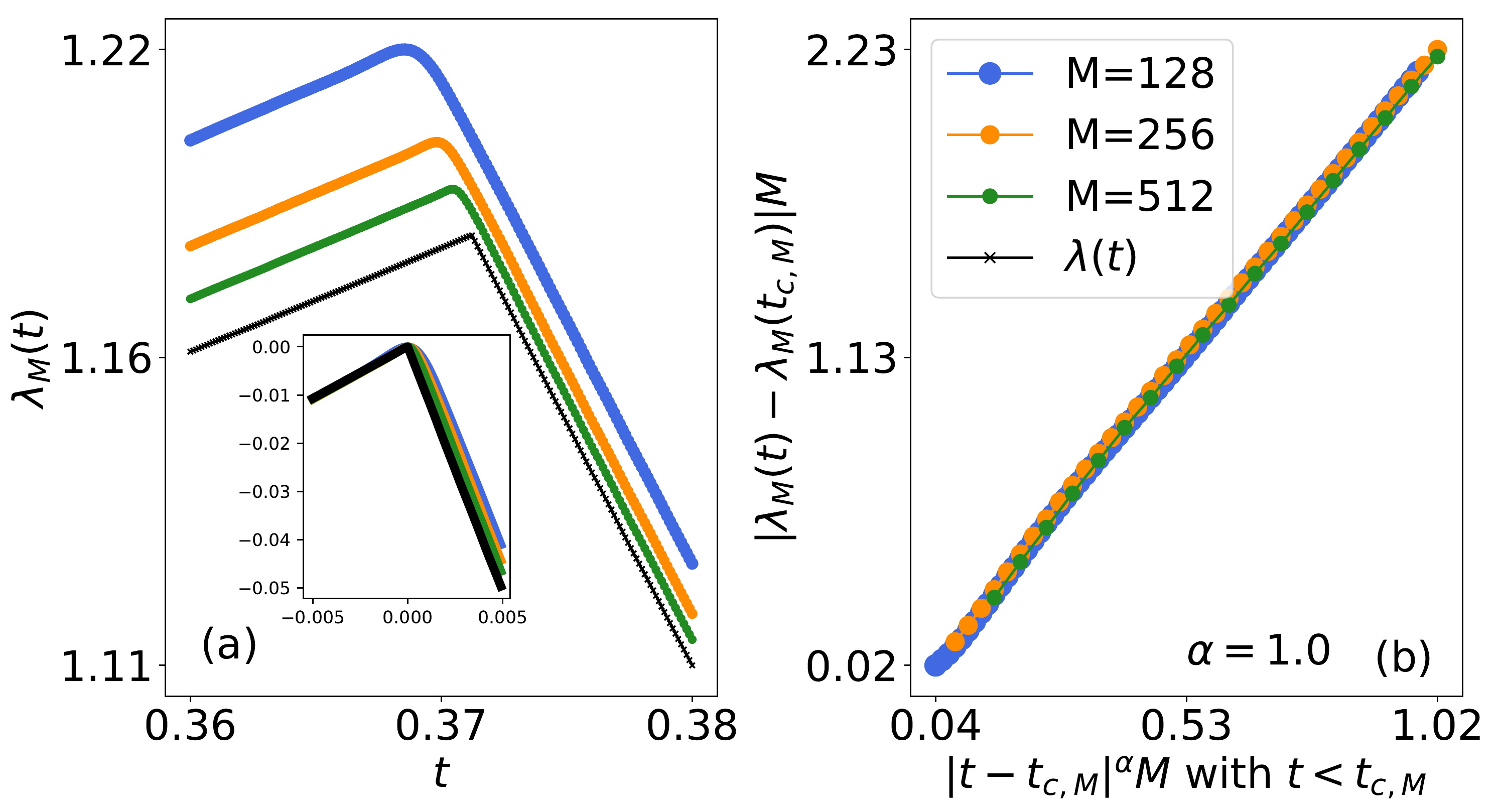}
	\caption{(Color online). Quench in the transverse-field strength of the long-range quantum Ising chain of Eq.~\eqref{H_long} for $\mu=1.8$ from $h_i=0$ to $h_f=5>h_c^d(\mu=1.8,h_i=0)\approx2.05$, which results in a regular DQPT at $t_c\sim0.371$. (a) Real-local effective free energy $\lambda_M(t)$ (colored dotted lines) as a function of time $t$ for three values of the configuration size $M=128,\,256,\,512$ sites, along with the exact effective free energy $\lambda(t)$ (black crossed line). The DQPT can be detected in $\lambda_M(t)$, where the corresponding peak sharpens with increasing $M$. The approximate critical time $t_{c,M}$ estimated from $\lambda_M(t)$ also becomes more accurate at larger $M$. The inset shows a zoom-in of $\lambda_M(t)-\lambda_M(t_{c,M})$ as a function of $t-t_{c,M}$, where the shifted real-local and exact effective free energies overlap nicely. (b) $\lvert\lambda_M(t) - \lambda_M(t_{c,M})\rvert M$ as a function of $\lvert t-t_{c,M}\rvert^\alpha M$. The value of the critical exponent $\alpha$ should be such that the curves exhibited in panel (a) collapse onto each other. The best overlap achieved occurs for $\alpha=1$.}
	\label{fig:ZQR1}
\end{figure}

\begin{figure}[ht!]
\centering
\includegraphics[width=1.0\columnwidth]{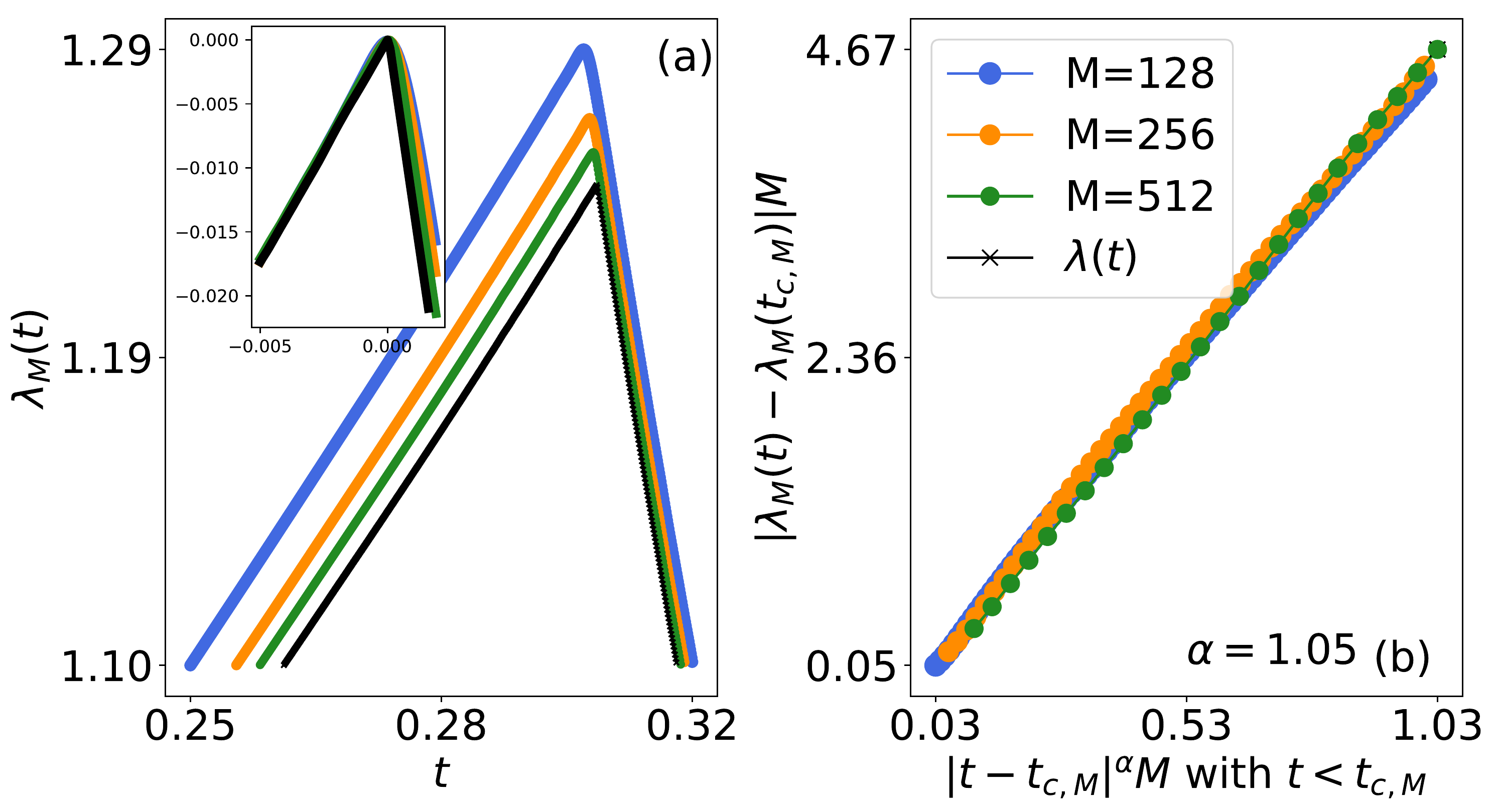}
\caption{(Color online). Quench in the transverse-field strength of the long-range quantum Ising chain of Eq.~\eqref{H_long} for $\mu=1.6$ from $h_i=0$ to $h_f=6>h_c^d(\mu=1.6,h_i=0)\approx2.35$, which results in a regular DQPT. (a) Real-local effective free energy $\lambda_M(t)$ (colored dotted lines) as a function of evolution time $t$ for three values of configuration size $M=128,\,256,\,512$ sites, in addition to the effective free energy $\lambda(t)$ (black crossed line). The DQPT can be detected in $\lambda_M(t)$ through the sharpening of the associated peak occurring at the estimated critical time $t_{c,M}$, which comes closer to the exact critical time $t_c$ with larger $M$. The inset shows a critical time $t_c\sim 0.3$, where we plot and zoom in on $\lambda_M(t)-\lambda_M(t_{c,M})$ as a function of $t-t_{c,M}$. The real-local and exact effective free energies fall nicely on top of each other, indicating good robustness of $\lambda_M(t)$ to capture this DQPT.  (b) $\lvert\lambda_M(t) - \lambda_M(t_{c,M})\rvert M$ as a function of $\lvert t-t_{c,M}\rvert^\alpha M$. The scaling analysis reveals the best collapse at a critical exponent $\alpha=1.05$.}
\label{fig:ZQR2}
\end{figure}

Let us now consider quenches starting in the ground state of Hamiltonian~\eqref{H_long} at initial transverse-field strength $h_i=0$ and quenching to $h_f>h_c^d(\mu,h_i)$. Quenches from the ordered phase to above the dynamical critical point $h_c^d(\mu,h_i)$ lead to regular cusps that generically correspond to zeros in the dynamics of the order parameter.\cite{zunkovic2016,halimeh2017dynamical} As an example, we set $\mu=1.8$ and $h_f=5>h_c^d(\mu=1.8,h_i=0)\approx2.05$. The resulting effective free energy and its real-local counterpart are shown in Fig.~\ref{fig:ZQR1}(a), where again we see that $\lambda_M(t)$ reliably discerns the DQPT and its associated critical time at large $M$, and even matches $\lambda(t)$ very well (see inset). The scaling analysis shown in Fig.~\ref{fig:ZQR1} yields the best collapse at a critical exponent $\alpha=1$, just as in the case of a regular DQPT when quenching from the paramagnetic to the ordered phase, which we have presented in Fig.~\ref{fig:regular}. Note here that the scaling function itself is roughly linear, differently from the quadratic ones of Fig.~\ref{fig:regular}(c,d) for the regular DQPTs due to a quench from the paramagnetic phase. As we will show later, this seems related to $\mu$ rather than the quench direction.

We repeat this analysis for $\mu=1.6$ while quenching from $h_i=0$ to $h_f=6>h_c^d(\mu=1.6,h_i=0)\approx2.35$. The performance of the real-local effective free energy, shown in Fig.~\ref{fig:ZQR2}(a), is qualitatively identical to that of Fig.~\ref{fig:ZQR1}(a). The critical time of the regular DQPT is approximated well by $t_{c,M}$, which approaches the actual critical time $t_c$ with larger $M$. The corresponding scaling analysis is shown in Fig.~\ref{fig:ZQR2}(b), where the best collapse seems to occur with a roughly linear scaling function at a critical exponent $\alpha=1.05$. This is slightly different from its counterpart in Fig.~\ref{fig:ZQR1}(b), which stands at unity. However, one has to be careful not to interpret too much into such a small difference, as the scaling analysis is not a highly accurate procedure. In principle $\alpha=1$ or $1.05$ cannot conclusively determine whether these two DQPTs are of different universality given the imperfect precision of our scaling analysis. Indeed, we have also considered different quenches that lead to regular DQPTs, and we find in all of them that the critical exponent is in the range $\alpha=1$--$1.1$; cf.~Appendix~\ref{sec:extra} for corresponding results and scaling analysis.

Nevertheless, we have two main take-home messages here. First, the real-local effective free energy $\lambda_M(t)$ is a robust detector of regular DQPTs independent of what phase the quench starts in and of the range of interactions. Second, the regular DQPTs exhibit universal scaling behavior even in a nonintegrable model such as the power-law interacting quantum Ising chain described by the Hamiltonian~\eqref{H_long}.

\subsubsection{Quenches from the ordered phase: anomalous DQPTs}\label{anomalous}
\begin{figure}[ht!]
	\centering
	\includegraphics[width=1.0\columnwidth]{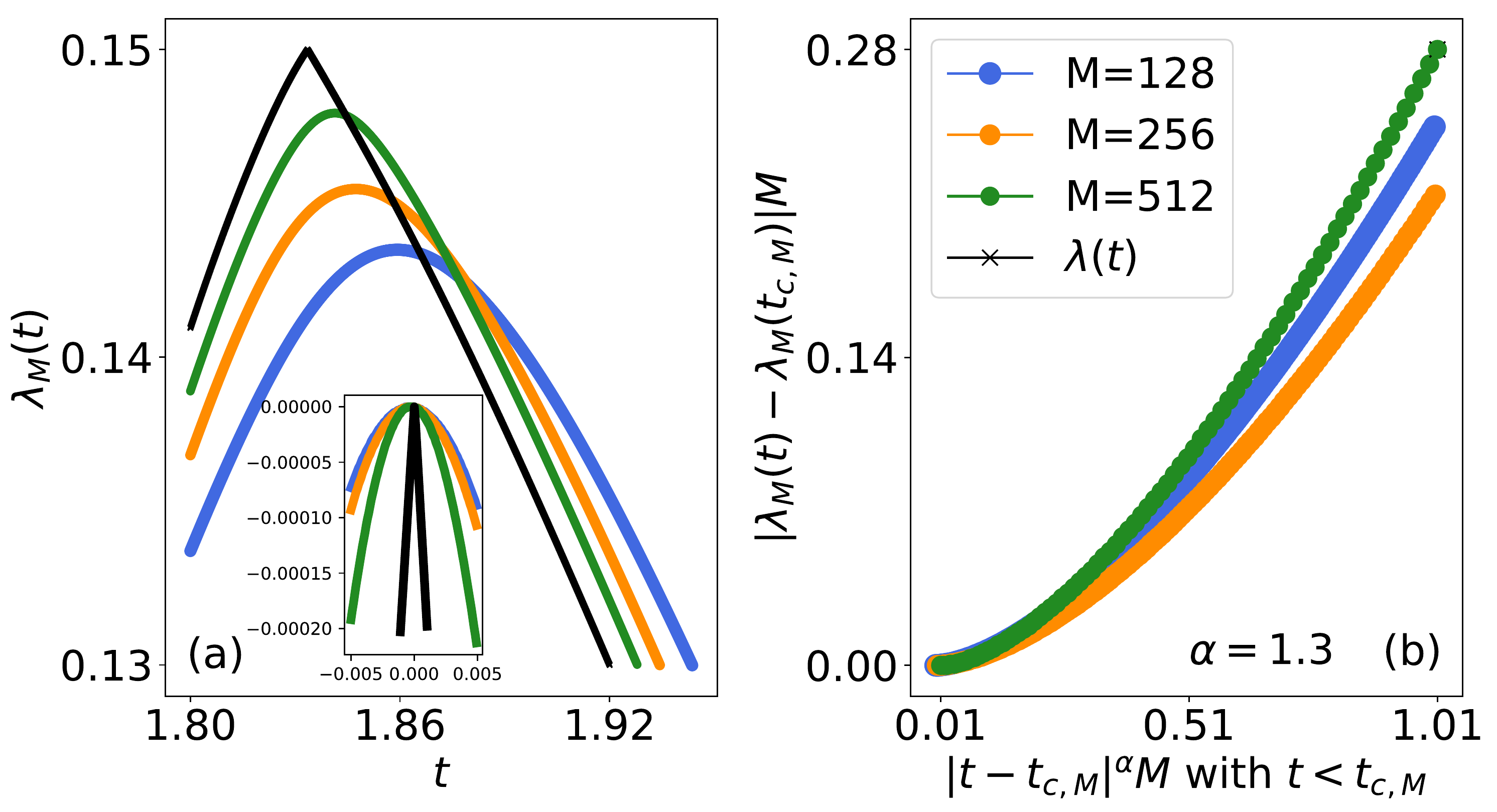}
	\caption{(Color online). Quench in the transverse-field strength of the long-range quantum Ising chain described by the Hamiltonian~\eqref{H_long} for $\mu=2$ from $h_i=0$ to $h_f=1.25<h_c^d(\mu=2)\approx1.85$, which results in an anomalous DQPT. (a) Real-local effective free energy $\lambda_M(t)$ (colored dotted lines) as function of time $t$ for three values of configuration size $M=128,\, 256,\,512$ sites, along with the exact effective free energy $\lambda(t)$ (black crossed line). Much the same way as in the case of regular DQPTs, here with increasing $M$ the real-local effective free energy $\lambda_M(t)$ exhibits a sharper peak and its estimated critical time $t_{c,M}$ approaches its exact counterpart $t_c\sim 1.84$ at which $\lambda(t)$ shows the anomalous DQPT. The inset displays a zoom-in of $\lambda_M(t)-\lambda_M(t_{c,M})$ as a function of $t-t_{c,M}$, indeed showing a sharpening peak with increasing $M$.  (b) $\lvert \lambda_M(t) - \lambda_M(t_{c,M})\rvert M$ as a function of $\lvert t-t_{c,M}\rvert^\alpha M$. The value of the critical exponent $\alpha$ should be such that the curves exhibited in panel (a) collapse onto each other. The best overlap achieved occurs for $\alpha=1.3$, although the result is conclusive only at relatively small $\lvert t-t_{c,M}\rvert$.}
	\label{fig:anomalous}
\end{figure}

Anomalous DQPTs occur for quenches within the ordered phase when low-lying quasiparticles in the spectrum of the quench Hamiltonian are local excitations, which in the case of one-dimensional Ising Hamiltonians amounts to domain-wall binding.\cite{Halimeh2018a} A suitable model for the observation of anomalous cusps is the Hamiltonian Eq.~\eqref{H_long} with $\mu=2$,\cite{halimeh2017dynamical} while quenching from $h_i=0$ to $h_f=1.25<h_c^d(\mu=2)\approx1.85$.

In Fig.~\ref{fig:anomalous}(a) we show the real-local effective free energy $\lambda_M(t)$ as a function of time for three different values of projective-configuration size $M=128,\,256,\,512$ sites, and the effective free energy $\lambda(t)$. We notice that this quantity exhibits a DQPT around $t_c \sim 1.84$, and with increasing $M$ the real-local effective free energy $\lambda_M(t)$ becomes sharper while the approximate critical time $t_{c,M}$ approaches its exact counterpart $t_c$ also. 

We now probe the universality of this anomalous DQPT through scaling analysis. The results are presented in Fig.~\ref{fig:anomalous}(b), which shows $\lvert\lambda_M(t) - \lambda_M(t_{c,M})\rvert M$ as a function of $\lvert t-t_{c,M}\rvert^\alpha M$. The critical exponent $\alpha$ should be chosen such that the curves for different values of $M$ collapse onto each other. The best overlap occurs for $\alpha =1.3$ although the result is not conclusive except for very small times. Therefore we cannot conclude that the real-local effective free energy $\lambda_M(t)$ is suitable to detect this anomalous DQPT. 

We have also checked other anomalous DQPTs for different values of $\mu$ and $h_f<h_c^d(\mu,h_i)$, and we also find that the best collapse occurs at $\alpha\approx1.2$--$1.3$. Nevertheless, since anomalous DQPTs occur at later times than their regular counterparts, the numerically accessible maximal bond dimension specified in iMPS is usually reached before the associated critical time, leading to higher inaccuracy in estimating the critical exponent $\alpha$. Nevertheless, one must also be open to the possibility that the collapse is less conclusive  for the anomalous DQPT of Fig.~\ref{fig:anomalous} simply because it may not be universal like its regular counterparts. This is indeed possible here because the exact and real-local effective free energies associated with this anomalus DQPT are completely converged with respect to maximal bond dimension (see Appendix~\ref{numspecs}). Further analysis of the universality of anomalous DQPTs is warranted, particularly in models that are numerically more tractable than the power-law interacting Ising chain. An ideal candidate model is the quantum Ising chain with exponentially decaying interactions, where anomalous cusps are known to arise.\cite{Halimeh2018a} We leave such a study, which is beyond the scope of the present paper, for an upcoming work.

\section{Momentum-local effective free energy $\lambda_k(t)$}\label{ML_lambda}
Next, we turn to systems where observations are more naturally made in momentum space, such as ultracold-atom implementations where time-of-flight measurements are a standard procedure.\cite{bloch2008many} Specifically, we will focus in the following on the Kitaev chain\cite{Kitaev2001} as a paradigmatic model for topological quantum matter, but with the addition of nearest-neighbor interactions at strength $U$, described by the Hamiltonian
\begin{align}\nonumber
		H =& -\sum_j\Big[\big(Jc^\dagger_{j} c_{j+1}+\Delta c_j^\dagger c^\dagger_{j+1}+\text{H.c.}\big)\\\label{H_kitaev}
		&+h\big(1-2c^\dagger_{j}c_j\big)+U c^\dagger_{j} c_{j} c^\dagger_{j+1} c_{j+1}\Big],
\end{align}
where $c_j$ is the fermionic annihilation operator on site $j$, obeying the canonical anticommutation relations $\{c_j,c_l\}=0$ and $\{c_j,c_l^\dagger\}=\delta_{j,l}$, and $J$ and $\Delta$ are the coupling and pairing constants, respectively. This \textit{interacting Kitaev chain} (IKC) is of particular interest in investigations of interaction effects on the stability of Majorana edge modes,\cite{Gangadharaiah2011,Rahmani2015a,Rahmani2015b} and has been shown to be experimentally realizable in Josephson junctions,\cite{Hassler2012} and also in optical lattices\cite{Pinheiro2013,Piraud2014} by mapping it onto the XYZ model in a magnetic field through a Jordan-Wigner transformation (see Appendix~\ref{mapping} for derivation and further details). In order to better understand the reasons lying at the basis of the definition of the momentum-local effective free energy $\lambda_k(t)$, which is suitable for this class of problems, it is useful to consider the dynamics emerging when a chain prepared in the ground state of Eq.~\eqref{H_kitaev} in the limit of $h\rightarrow \infty$ is subsequently quenched by the same model at a finite $h$ in the noninteracting limit $U=0$. 
Our interest in such a quench protocol resides in the fact that the resulting effective free energy $\lambda(t)$ can be derived analytically, and will serve as the starting point for the introduction of the momentum-local return probability $\lambda_k(t)$.
As such, we set $U=0$ in Eq.~\eqref{H_kitaev} and employ the Fourier transformation
\begin{align}\label{FT}
        c_j = \frac{1}{\sqrt{N}} \sum_{k}^\text{B.z.} e^{ikj} c_k,
\end{align}
where B.z.~denotes the Brillouin zone $[-\pi,\pi)$, leading to
\begingroup
\renewcommand{\arraystretch}{1.5}
\begin{align}\label{eq:BdG}
H&=-\sum_k^\text{B.z.}\Psi_k^\dagger H_k\Psi_k,\\\nonumber
\Psi_k&=\begin{pmatrix}
c_k \\
c_{-k}^\dagger
\end{pmatrix},\,\,\,H_k=\begin{pmatrix}
h-J\cos k & -i\Delta\sin k \\
i\Delta\sin k & J\cos k-h
\end{pmatrix}.
\end{align}
\endgroup
The Bogoliubov-de Gennes Hamiltonian of Eq.~\eqref{eq:BdG} can be diagonalized by a Bogoliubov transformation
\begin{subequations}\label{bogo}
\begin{align}
c_k&=\cos\theta_k\,\gamma_k - i\sin\theta_k\,\gamma^{\dagger}_{-k}, \\
c^{\dagger}_{-k}&=\cos\theta_k\,\gamma^{\dagger}_{-k}-i\sin\theta_k\,\gamma_k,\\\label{eq:thetak}
\theta_k&=\frac{1}{2}\arctan\bigg(\frac{\Delta \sin k}{h-J\cos k}\bigg).
\end{align}
\end{subequations}
where $\gamma_k^{(\dagger)}$ are Bogoliubov fermionic operators with the canonical anticommutation relations $\{\gamma_j,\gamma_l\}=0$ and $\{\gamma_j,\gamma_l^\dagger\}=\delta_{j,l}$. This allows us to rewrite Eq.~\eqref{H_kitaev} at $U=0$ in the diagonal form
\begin{subequations}\label{H_d}
\begin{align}
	H &= \sum_{k>0} E_k \big( \gamma_k^{\dagger}\gamma_k - \gamma_{-k}\gamma^{\dagger}_{-k}  \big),\\
    E_k &= \sqrt{(h-J\cos k)^2 + \Delta^2\sin^2 k}.
\end{align}
\end{subequations}
We note here that the Bogoliubov operators $\gamma_k$ are not the same pre- and post-quench, corresponding to $h=h_i$ and $h=h_f$, respectively, since for each value of $h$, there is generically a unique set of Bogoliubov operators  $\gamma_k$ that diagonalize the Hamiltonian~\eqref{eq:BdG}. Following the approach of BCS theory,\cite{Bardeen1957} the ground state of Eq.~\eqref{H_d} at $h=h_i$ in terms of the \textit{post-quench} Bogoliubov operators reads
\begin{align}\nonumber
	\ket{\text{GS}} &= \frac{1}{\mathcal{N}} \exp\bigg(i \sum_{k>0} \Lambda_k \gamma^{\dagger}_{k}\gamma^{\dagger}_{-k}\bigg) \ket{0} \\\label{GS}
	&=  \frac{1}{\mathcal{N}}  \prod_{k>0} \Big(1+ i\Lambda_k \gamma^{\dagger}_{k}\gamma^{\dagger}_{-k}\Big) \ket{0},
\end{align}
where $\ket{0}$ is the vacuum of the \textit{post-quench} fermionic operator $\gamma^{\dagger}_k$, and $\mathcal{N}$ is a normalization factor given by
\begin{subequations}
\begin{align}\label{normalization}
	\mathcal{N}^2 &= \prod_{k>0} \mathcal{N}_k^2 = \prod_{k>0} \left[1+ \Lambda_k^2 \right],\\
	\Lambda_k &= \tan(\theta_k^f-\theta_k^i),
\end{align}
\end{subequations}
where the superscript $i$ ($f$) corresponds to $h=h_i$ ($h=h_f$) in Eq.~\eqref{eq:thetak}.

Applying the unitary time evolution operator $e^{-iHt}$, with $H$ as given in Eq.~\eqref{H_d} at $h=h_f$, to the initial state of Eq.~\eqref{GS}, we obtain for the time-evolved state at a generic time $t$:
\begin{equation}
\begin{split}
\ket{\psi(t)} = \frac{1}{\mathcal{N}}  \prod_{k>0} \Big[1+ ie^{-2iE_kt}\Lambda_k \gamma^{\dagger}_{k}\gamma^{\dagger}_{-k}\Big]  \ket{0}.
\end{split}
\label{psi}
\end{equation}
The Loschmidt amplitude $\mathcal{L}_{U=0}(t)$ is momentum-factorizable in the noninteracting case we consider here, and as such, it can be written as
\begin{align}\nonumber 
		\mathcal{L}_{U=0}(t) &= \bra{\psi(0)}\ket{\psi(t)} = \frac{1}{\mathcal{N}^2} \prod_{k>0} \bra{\psi_k(0)}\ket{\psi_k(t)}\\\label{LA}
		&= \frac{1}{\mathcal{N}^2} \prod_{k>0}  \left[ 1 + \Lambda_k^2 e^{-2iE_kt} \right],
\end{align}
where $\ket{\psi_k(t)}=  \big(1+ ie^{-2iE_kt}\Lambda_k \gamma^{\dagger}_{k}\gamma^{\dagger}_{-k}\big)\ket{0}.$
The Loschmidt amplitude can be expressed in terms of momentum uncorrelated two-body observables at the same time $t$. The final result yields
\begin{figure}[ht!]
	\centering
	\includegraphics[width=\columnwidth]{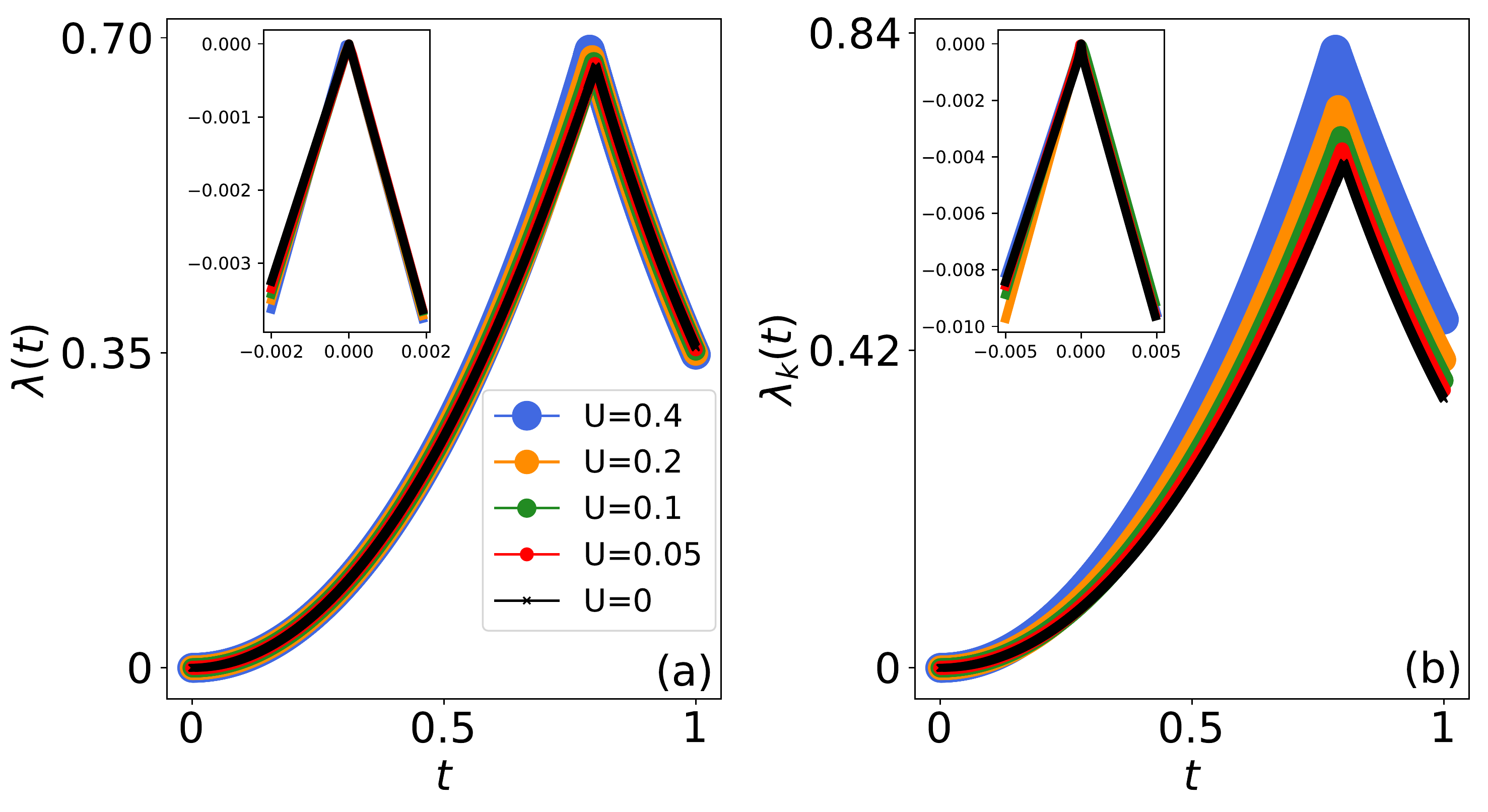}
	\caption{(Color online). Dynamics of the interacting Kitaev chain given by Eq.~\eqref{H_kitaev} for $J=\Delta=1$, and upon quenching $h$ from $h_i\to\infty$ to $h_f=0.2$. (a) Effective free energy $\lambda(t)$ as a function of time $t$ for different values of $U=0,\,0.05,\,0.1,\,0.2,\,0.4$. For each of these values $\lambda(t)$ shows a cusp around $t_c \sim 0.8$. The inset shows a zoom-in of $\lambda(t)-\lambda(t_c)$ as a function of $t-t_c$ around $t-t_c\approx0$. (b) Momentum-local effective free energy $\lambda_k(t)$ as a function of evolution time $t$ for the same values of the interaction strength $U$ considered in panel (a). For $U=0$, we have the exact limit $\lambda_k(t)=\lambda(t)$. The momentum-local effective free energy reliably detects the underlying DQPT exhibiting a sharp feature around $\tilde{t}_c \sim 0.8$, where we see that the approximate critical time $\tilde{t}_c$ extracted from $\lambda_k(t)$ approaches $t_c$ with decreasing $U$. The inset shows a zoom-in of $\lambda_k(t)-\lambda_k(\tilde{t}_c)$ as a function of $t-\tilde{t}_c$ while zooming in around $t-\tilde{t}_c\approx0$.}
	\label{fig:IKC}
\end{figure}

\begin{align}\label{LA_prod}
  \mathcal{L}_{U=0}(t) %= \frac{1}{\mathcal{N}^2} \prod_{k>0} \left[ 1 + \left( \frac{V_k}{U_k} \right)^2 e^{-i2tE_k} \right]   \\   
  = \prod_{k>0} \frac{d_k(0) d_k(t) - b_k(0) b_k(t)}{ d_k(0) d_k(0) - b_k(0) b_k(0)},
\end{align}
where
\begin{align}
	\label{scal_prod_main}
	b_k(t) &= \bra{\psi(t)}\gamma_k \gamma_{-k}\ket{\psi(t)} =  \frac{i}{\mathcal{N}_k^2}\Lambda_k e^{-2iE_kt},\\
	\label{numb_inv_main}
d_k(t)& = \bra{\psi(t)}\gamma_{k} \gamma^{\dagger}_{k}\ket{\psi(t)}=\frac{1}{1+\Lambda_k^2} = \frac{1}{\mathcal{N}_k^2}.
\end{align}

As mentioned, one of the advantages of the form of the Loschmidt amplitude in Eq.~\eqref{LA_prod} is in its factorization in terms of two-body observables, which are local measures from an experimental perspective. Defining the momentum-local effective free energy
\begin{align}
\lambda_k(t)=-\lim_{N\to\infty}\frac{1}{N}\ln\lvert\mathcal{L}_{U=0}(t)\rvert^2,
\end{align}
with $\mathcal{L}_{U=0}(t)$ given by Eq.~\eqref{LA_prod}, we see that whereas when $U=0$ it is identically the effective free energy $\lambda(t)$ of the model in Eq.~\eqref{H_kitaev}, for $U\neq0$ when interactions are on $\lambda_k(t)$ becomes an approximation for $\lambda(t)$, because in this case the expectation values of the quadratic terms in Eqs.~\eqref{scal_prod_main} and~\eqref{numb_inv_main} provide a mean-field type of approximation for higher-order terms that generically contribute to the effective free energy in the presence of interactions. As we demonstrate in what follows, this allows the use of $\lambda_k(t)$ as a reliable measure of DQPTs that is \textit{local} in momentum space.

We prepare our system in the ground state of the IKC at $h_i\to\infty$. This corresponds to an empty chain in the physical space of the fermionic operators $c_j$. We then quench to $h_f=0.2$, with $J=\Delta=1$, while setting the interaction strength to one of several values $U=0,\,0.05,\,0.1,\,0.2,\,0.4$. The ensuing dynamics of the effective free energy $\lambda(t)$ and its momentum-local approximation $\lambda_k(t)$ are shown in Fig.~\ref{fig:IKC}(a,b), respectively. Interestingly, we see that $\lambda_k(t)$ reliably captures the sharp feature of a DQPT even when $U=0.4$. Let us denote by $t_c$ the exact critical time at the DQPT arising in $\lambda(t)$, while calling its approximate counterpart from $\lambda_k(t)$ as $\tilde{t}_c$. As shown by comparing the insets that zoom-in on $\lambda(t)-\lambda(t_c)$ as a function of  $t-t_c$  around $t\approx t_c$ in Fig.~\ref{fig:IKC}(a), and on $\lambda_k(t)-\lambda_k(\tilde{t}_c)$ as a function of $t-\tilde{t}_c$ around $t\approx \tilde{t}_c$ in Fig.~\ref{fig:IKC}(b), the feature of a sharp DQPT is reproduced reliably in the momentum-local approximation. Moreover, the estimated critical times $\tilde{t}_c$ reliably approximate their exact counterparts $t_c$, where at small $U$ they are roughly identical, with both shifting to the left with large $U$. However, nonuniversal features, such as amplitudes of the effective free energies, are not as robustly approximated. Overall, this bodes well for experimental efforts focused on detecting DQPTs in momentum space in models of topological quantum matter with added interactions.
	
\section{Concluding discussion} \label{conclusions}
In this work we have introduced two local measures for the reliable and experimentally feasible observation of DQPTs. Whereas the exact effective free energy is a global quantity, its \textit{real-local} and \textit{momentum-local} counterparts involve a projection of the time-evolved wave function onto a finite configuration in real space and two-point correlations in momentum space, respectively. These measures can be beneficial in modern ultracold-atom and ion-trap experiments in that they allow for robust detection of DQPTs with exponentially fewer resources than in the case where measurement of the exact effective free energy is attempted. We have demonstrated the efficacy of these measures on several paradigmatic models. 

The real-local effective free energy efficiently captures DQPTs in systems with degrees of freedom well-defined in real space, such as spin models. We have tested this measure there in a quench involving the nearest-neighbor Ising model, in addition to various quenches in the quantum Ising chain with power-law decaying interactions. In various cases, the real-local effective free energy proves a reliable tool for discerning universal behavior and extracting associated critical exponents of DQPTs. Indeed, scaling analysis through this local measure indicates that anomalous and regular DQPTs exhibit significantly different dynamical criticality. Even though scaling analysis produces strong evidence of universal behavior in regular DQPTs with a critical exponent $\alpha\approx1$--$1.1$, such a conclusion is less clear when it comes to anomalous DQPTs. Notwithstanding this difference, the real-local effective free energy shows impressive reliability in detecting both types of DQPTs and approximates their critical times accurately.

The momentum-local effective free energy is demonstrated on the interacting Kitaev chain, which is a stability testbed for Majorana edge modes in the presence of interactions. This local measure, exact in the noninteracting limit, reliably captures DQPTs at finite $U$. This is a remarkable result, because in interacting fermionic systems the effective free energy generically involves arbitrarily high orders of correlations in momentum space, because the system cannot be decomposed into disconnected momentum sectors. Nevertheless, the momentum-local effective free energy, which involves only two-point correlations defined by a single momentum value, even reliably captures the critical time of the DQPT at finite $U$, albeit the difference between the captured critical time and the actual one increases with $U$.

\begin{acknowledgements}
This project has received funding from the European Research Council (ERC) under the European Unions Horizon 2020 research and innovation programme (grant agreement No. 853443), and M.H.~further acknowledges support by the Deutsche Forschungsgemeinschaft (DFG) via the Gottfried Wilhelm Leibniz Prize program. The authors are grateful to Ian P.~McCulloch for stimulating discussions. J.C.H.~acknowledges support by the Interdisciplinary Center Q@TN --- Quantum Science and Technologies at Trento, the DFG Collaborative Research Centre SFB 1225 (ISOQUANT), the Provincia Autonoma di Trento, and the ERC Starting Grant StrEnQTh (Project-ID 804305).
\end{acknowledgements}

\appendix

\section{Further results on the long-range quantum Ising chain}\label{sec:extra}

\begin{figure}[ht!]
	\centering
	\includegraphics[width=1.0\columnwidth]{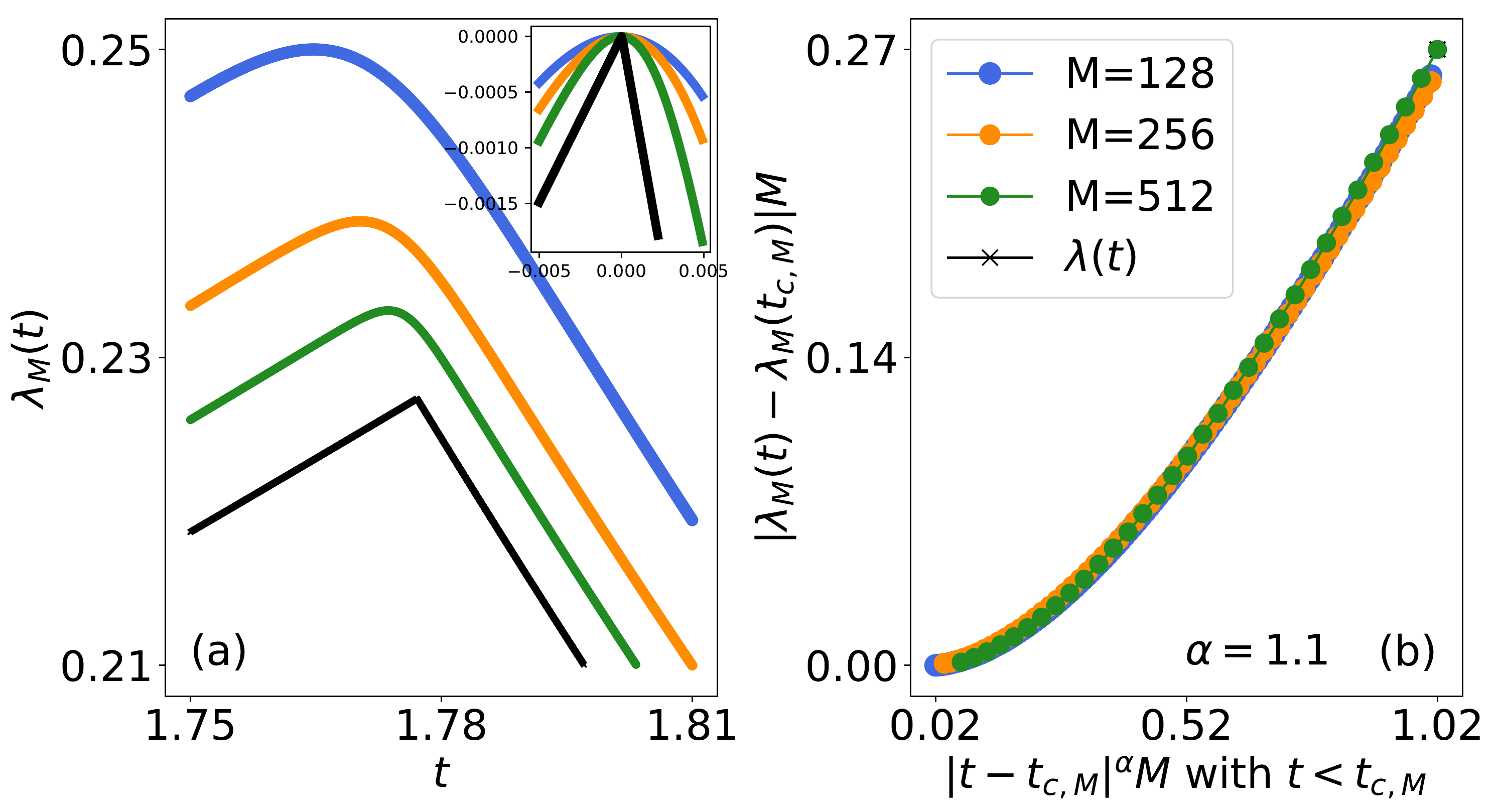}
	\caption{(Color online). Dynamics in the long-range quantum Ising chain given by Eq.~\eqref{H_long} with $\mu=2$, in the wake of a quench in the transverse-field strength from $h_i=0$ to $h_f=3>h_c^d(\mu=2,h_i=0)\approx1.85$, leading to a regular DQPT. (a) Effective free energy (black crossed line) $\lambda(t)$ and its real-local counterpart (colored dotted lines) $\lambda_M(t)$ for configuration size $M=128,\,256,\,512$ sites as function of time $t$. Whereas $\lambda(t)$ shows a clear DQPT, $\lambda_M(t)$ is smooth, although we see a sharpening of the associated peak with increasing $M$, as well as better accuracy in estimating the approximate critical time $t_{c,M}$ in the associated peak of $\lambda_M(t)$. The inset shows a zoom-in of $\lambda_M(t)-\lambda_M(t_{c,M})$ as a function of $t-t_{c,M}$. (b) Scaling analysis where we show $\lvert\lambda_M(t)-\lambda_M(t_{c,M})\rvert M$ as a function of $\lvert t-t_{c,M}\rvert^\alpha M$. The best collapse occurs for $\alpha=1.1$, and seems rather conclusive.}
	\label{fig:ZQR1_Appendix}
\end{figure}

\begin{figure}[ht!]
	\centering
	\includegraphics[width=1.0\columnwidth]{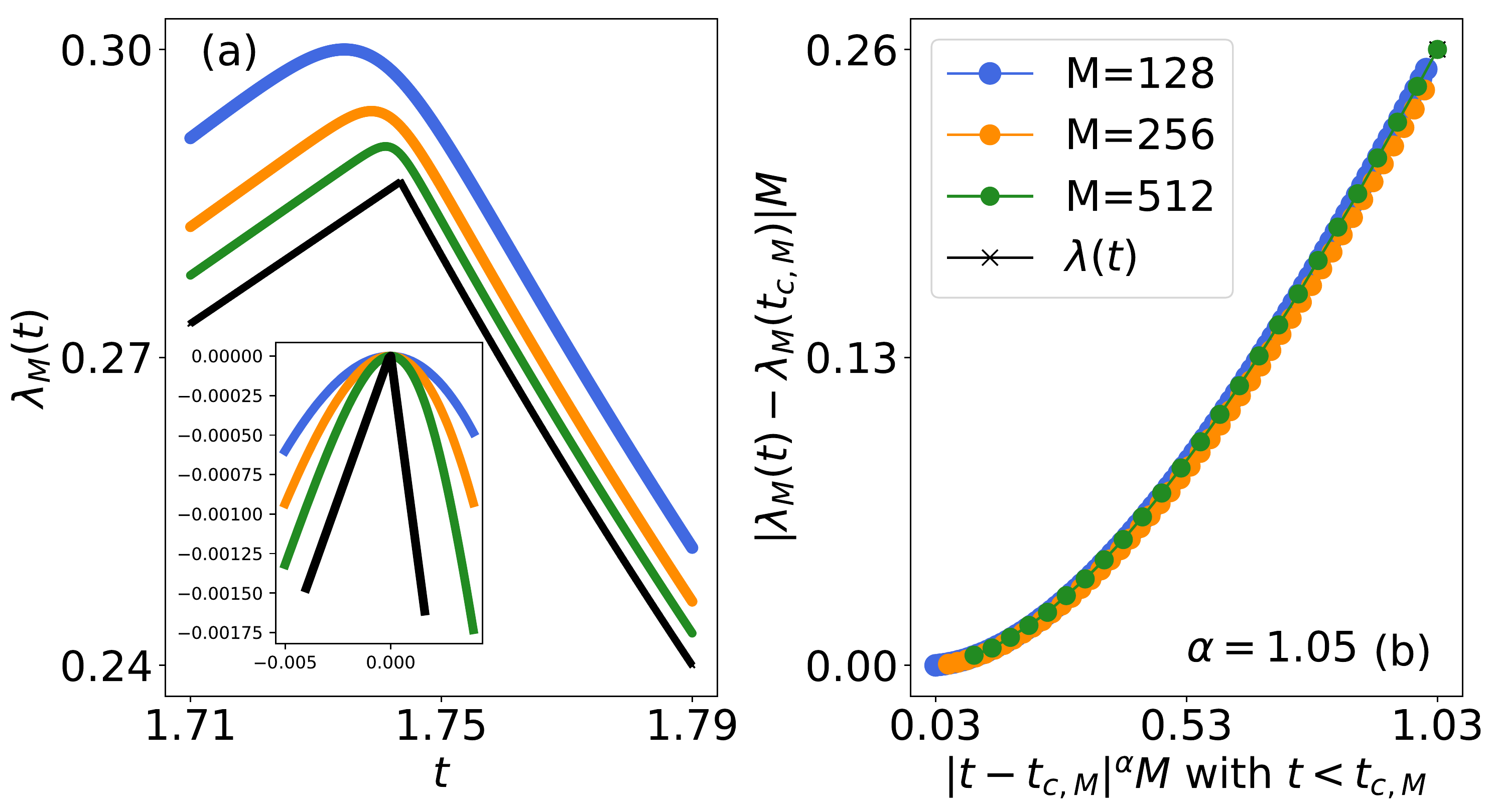}
	\caption{(Color online). Same as Fig.~\ref{fig:ZQR1_Appendix} but for $\mu=2.2$, the dynamical critical point of which is $h_c^d(\mu=2.2,h_i=0)\approx1.7$. Unlike Fig.~\ref{fig:ZQR1_Appendix}, the critical exponent that gives the best collapse in the scaling analysis is $\alpha=1.05$ rather than $1.1$. Nevertheless, within the precision of our scaling analysis, one cannot conclude that the associated DQPTs are of different universality.}
	\label{fig:ZQR2_Appendix}
\end{figure}

\begin{figure}[ht!]
	\centering
	\includegraphics[width=1.0\columnwidth]{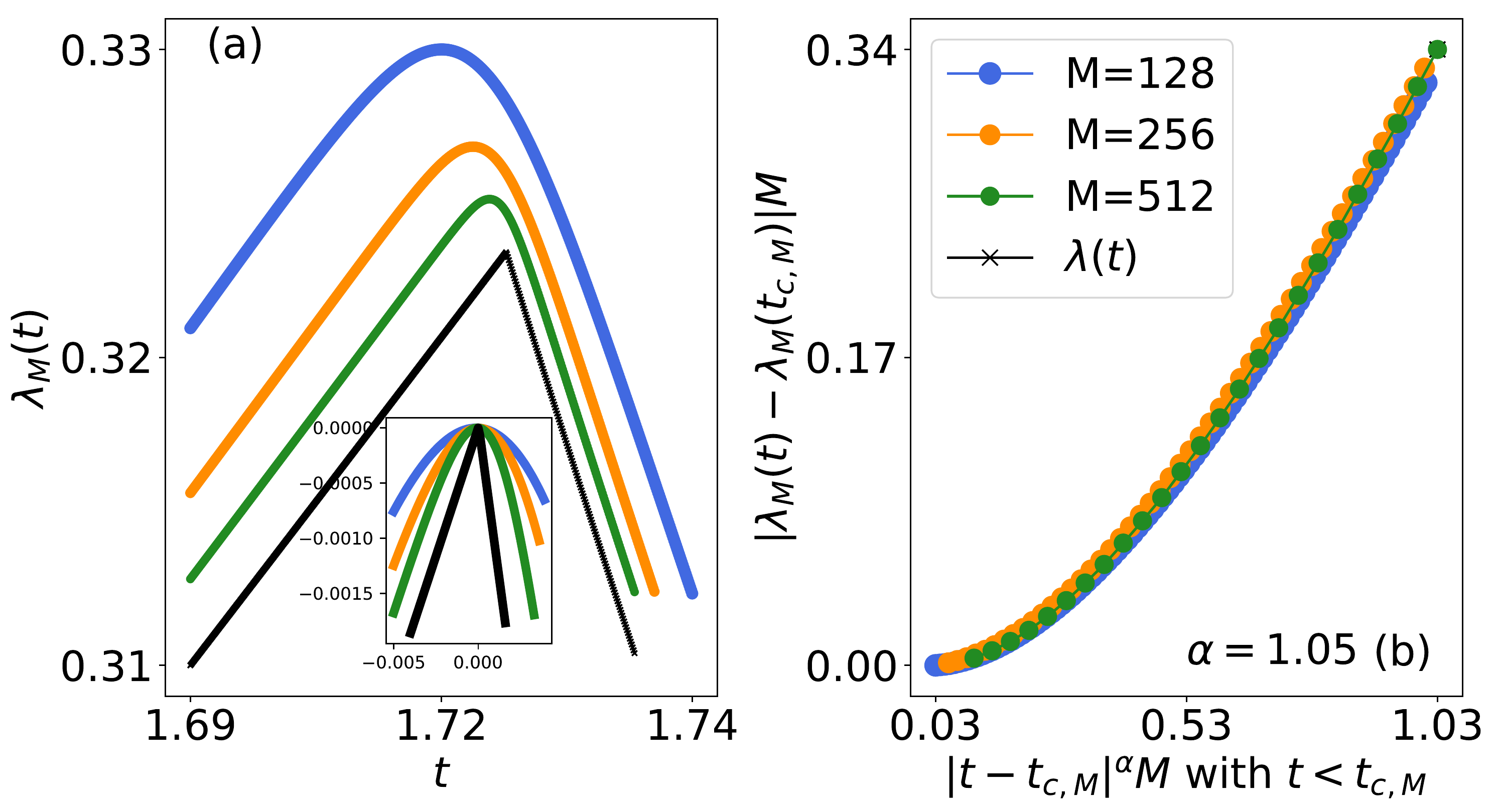}
	\caption{(Color online). Same as Figs.~\ref{fig:ZQR1_Appendix} and~\ref{fig:ZQR2_Appendix} but for $\mu=2.4$, the dynamical critical point of which is $h_c^d(\mu=2.4,h_i=0)\approx1.51$. The scaling analysis yields the best collapse for $\alpha=1.05$. The precision of our scaling analysis cannot definitively ascertain whether this regular DQPT is of a different universality than its counterparts in Figs.~\ref{fig:ZQR1} and~\ref{fig:ZQR1_Appendix}. }
	\label{fig:ZQR3_Appendix}
\end{figure}

In the main text, we have presented scaling-analysis results on various DQPTs in the long-range transverse-field Ising chain of Eq.~\eqref{H_long} for quenches from initial value $h_i$ of the transverse-field strength to a final value of $h_f$. The aforementioned results suggest that regular DQPTs, which occur for quenches from $h_i>h_c^e(\mu)$ to $h_f<h_c^e(\mu)$ or from $h_i<h_c^e$ to $h_f>h_c^d(\mu,h_i)$, exhibit a critical exponent $\alpha=1$--$1.05$. On the other hand, anomalous DQPTs, which occur in quenches $h_i<h_c^e$ to $h_f<h_c^d(\mu,h_i)$ when the quench Hamiltonian at $h_f$ hosts domain-wall binding in its spectrum, seem to have a different critical exponent $\alpha=1.2$--$1.3$, although it is possible that they may simply not be universal. 

Here, we add results for three further regular DQPTs for the same quench in the long-range quantum Ising chain from $h_i=0$ to $h_f=3>h_c^d(\mu,h_i=0)$ for $\mu=2,\,2.2,\,2.4$. Figure~\ref{fig:ZQR1_Appendix}(a) shows the effective free energy and its real-local counterpart $\lambda_M(t)$ for several values of the configuration size $M=128,\,256,\,512$ sites for the long-range quantum Ising chain at $\mu=2$, where $h_c^d(\mu=2,h_i=0)\approx1.85$. As in the results of the main text for regular DQPTs, $\lambda_M(t)$ shows a sharper peak at the approximate critical time $t_{c,M}$ with increasing $M$ (see inset), while $t_{c,M}$ also approaches the exact critical time $t_c$. A scaling analysis is carried out in Fig.~\ref{fig:ZQR1_Appendix}(b), where we find the best collapse to occur at $\alpha=1.1$. This is different from the values we get for this critical exponent for the regular DQPTs of Secs.~\ref{NN} and~\ref{long}, but within the precision of our scaling analysis, still not different enough to conclusively rule that this DQPT is of different universality. Note how unlike the cases of the regular DQPTs for $h_i=0$ in the main text, the scaling function for the regular DQPT in Fig.~\ref{fig:ZQR1_Appendix}(b) is nonlinear.

We repeat this analysis for two more values of $\mu=2.2$ and $2.4$ in Figs.~\ref{fig:ZQR2_Appendix} and~\ref{fig:ZQR3_Appendix}, respectively. The respective dynamical critical points for the quantum Ising Hamiltonian~\eqref{H_long} at these interaction ranges are $h_c^d(\mu=2.2,h_i=0)\approx1.7$ and $h_c^d(\mu=2.4,h_i=0)\approx1.51$. Much the same way as in the case of other regular DQPTs considered in this work,  with increasing configuration size $M$, the real-local effective free energy approximates its exact counterpart well, with a sharpening peak $\lambda_M(t_{c,M})$ at the approximate critical time $t_{c,M}$, which in turn approaches $t_c$; cf.~Figs.~\ref{fig:ZQR2_Appendix}(a) and~\ref{fig:ZQR3_Appendix}(a) and corresponding insets. The associated scaling analyses are shown in Figs.~\ref{fig:ZQR2_Appendix}(b) and~\ref{fig:ZQR3_Appendix}(b), where in both cases we find that the best collapse is achieved at $\alpha=1.05$ with a nonlinear scaling function, just as in the case of Fig.~\ref{fig:ZQR1_Appendix}(b). Also here, we note that due to the absence of indefinite precision in our scaling-analysis procedure, we cannot rule out that all the regular DQPTs we have analyzed are in reality of the same universality with an equal fixed value of the critical exponent $\alpha$ that lies in the range $[1,1.1]$, the exact determination of which is beyond our reasonable numerical capabilities.

The conclusion drawn from the results of this Appendix mirrors that of the main text. The real-local effective free energy $\lambda_M(t)$ is an impressive tool that reliably detects the presence of DQPTs and their corresponding critical times. Even more, in the case of regular DQPTs, it admits a scaling analysis that reveals universal behavior. These capabilities would be of benefit to modern ultracold-atom and ion-trap experiments attempting to detect nonanalytic behavior in the effective free energy.

\section{Numerical specifications}\label{numspecs}
We provide here further details that render feasible our iMPS implementation of the models and corresponding quenches considered in this work. Additionally, we also discuss convergence with respect to iMPS parameters, and provide convergence results for our most computationally demanding calculations.

\subsection{Long-range Hamiltonians}
The power-law decaying interactions in the Hamiltonian~\eqref{H_long} are not possible to implement exactly in the framework of iMPS, because traditionally the latter is based on matrix product operator (MPO) descriptions of exponentials of Hamiltonian parts containing only commuting parts.\cite{Pirvu2010,Uli2011} This works well for systems with short-range interactions since a finite MPO description is then possible, but is generically not possible for long-range interactions. However, a workaround exists involving the MPO representation of the interaction term in Eq.~\eqref{H_long}, as a sum of MPO representations of exponentials whose sum faithfully approximates the power-law interaction profile.\cite{Crosswhite2008} This approximation takes the form
\begin{align}
\sum_{m<n}\lvert m-n\rvert^{-\mu}\sigma^x_m\sigma^x_n\approx \sum_{m<n}\sum_{r=1}^Lc_ru_r^{\lvert m-n\rvert-1},
\end{align}
where $c_r\in\mathbb{R}$, $u_r\in[0,1)$, and $L$ is the number of exponentials used in the approximation. The cofficients $c_r$ and $u_r$ are computed through a nonlinear least-squares fit, with $L$ chosen appropriately over a distance $d$ large enough such that $\sum_{r=1}^Lc_ru_r^{d-1}<\varepsilon$. In our calculations we have set $\varepsilon\approx\mathcal{O}(10^{-8})$, which amounts to $L$ in the range of $5$--$25$ depending on the value of $\mu$.

As for the time evolution itself, this was carried out through integrating, in the thermodynamic limit, the Schr\"{o}dinger equation by applying TDVP on the MPS variational manifold. A Lie-Trotter scheme is employed for splitting the projector on the tangent space of the variational manifold, allowing the direct integration of the MPS-tensor effective differential equations. This is distinct from conventional Lie-Trotter splitting schemes applied on the global time-evolution operator. For a detailed description of the implementation, we refer the reader to Refs.~\onlinecite{Haegeman2011,Haegeman2016}. This splitting scheme works only for sufficiently small time-steps. In our numerical calculations we have used time-steps as small as $\tau=10^{-4}$, where for convenience we have set the energy scale $J=1$. This is far smaller than the required value for convergence ($\tau_\text{conv}\approx0.002$--$0.005$), but has nevertheless been necessary in terms of a reliable scaling analysis around the critical times of the effective free energy. An indefinitely precise scaling analysis would in principle require $\tau\to0$, albeit this is impractical, and within our computational resources, $\tau=10^{-4}$ is ideal at the cost of an additional contribution to the imprecision in determining the exact critical exponent $\alpha$ due to a noninfinitesimal time-step.

\begin{figure}[ht!]
	\centering
	\includegraphics[width=1.0\columnwidth]{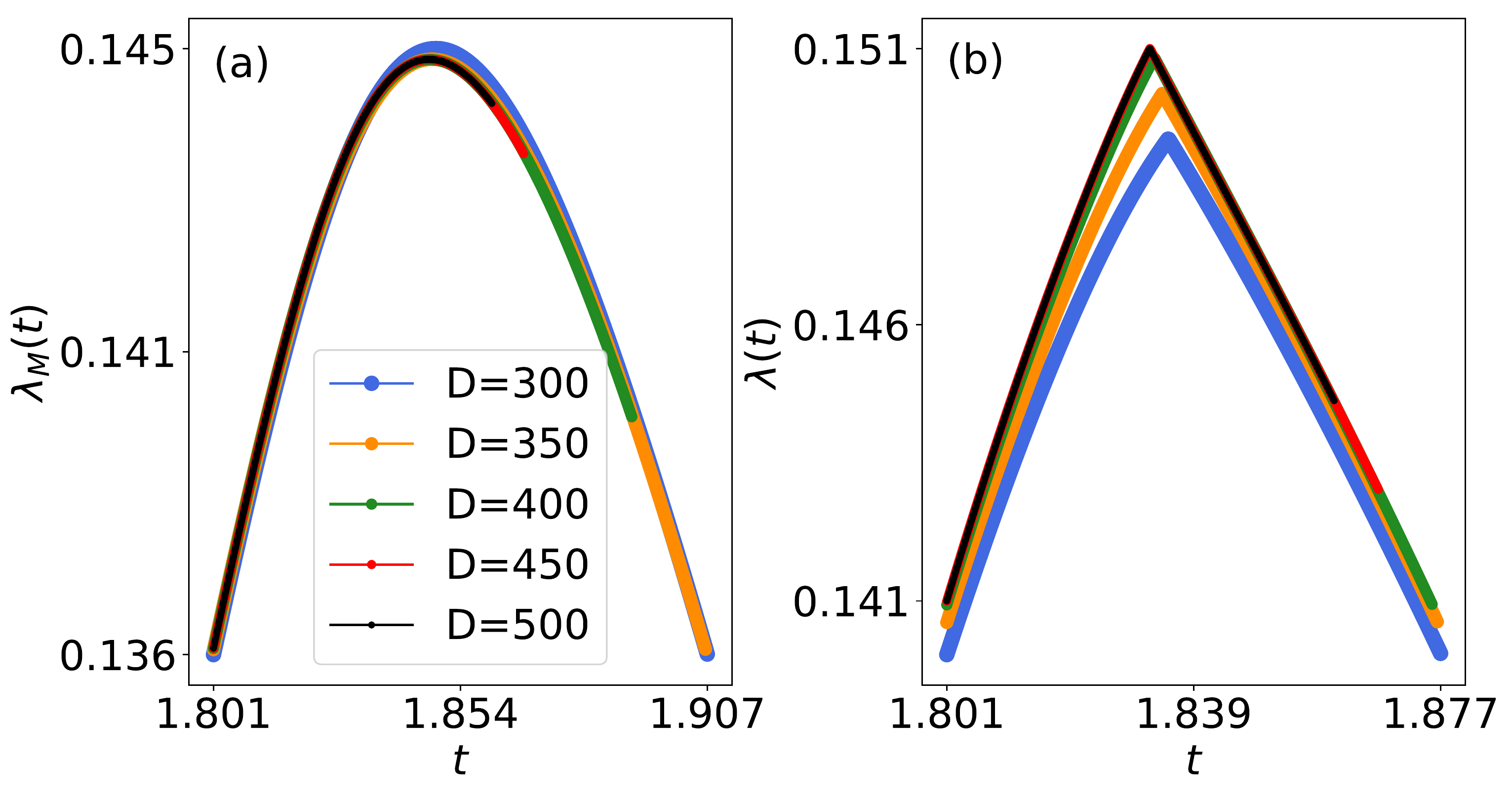}
	\caption{(Color online). Convergence with MPS maximal bond dimension $D$ in the case of the anomalous DQPT discussed in Sec.~\ref{anomalous} and shown in Fig.~\ref{fig:anomalous}, which is one of the most computationally demanding of our calculations. (a) The real-local effective free energy at configuration size $M=512$ sites shows good convergence already at maximal bond dimension $D=400$. (b) The exact effective free energy shows good convergence already at maximal bond dimension $D=450$.}
	\label{fig:convergence}
\end{figure}

Another iMPS knob to control for is the \textit{maximal bond dimension} $D$. This severely limits the maximal evolution time reached in an iMPS calculation due to the linear growth of entanglement entropy in case of generic global quenches, which is the case in our study. One of the most demanding calculations in this study was that of the anomalous DQPT in Sec.~\ref{anomalous}. These DQPTs generically occur after several smooth cycles in the effective free energy, and are thus usually delayed in time with respect to their regular counterparts, thereby requiring more computational resources to reliably capture them. We present in Fig.~\ref{fig:convergence} convergence results, where the real-local effective free energy $\lambda_{M=512}(t)$ shows good convergence at a maximal bond dimension $D=400$, while the exact effective free energy requires a maximal bond dimension of $D=450$. Just as in a laboratory experiment, $\lambda(t)$ is also computationally more expensive than its real-local approximation.

\subsection{Fermionic Hamiltonians}
Fermionic models like the IKC Hamiltonian~\eqref{H_kitaev} are cumbersome to implement in iMPS due to the fermionic anticommutation relations, which ideally we want to avoid. One way of achieving this is to employ a mapping onto spin or bosonic systems. In our iMPS calculations, we implement the IKC Hamiltonian~\eqref{H_kitaev} by mapping it onto the XYZ model in a magnetic field, given in Eq.~\eqref{H_XYZ}.

\section{Mapping the IKC to the XYZ model in a magnetic field}\label{mapping}
The IKC Hamiltonian~\eqref{H_kitaev} can be mapped onto a spin model through the Jordan-Wigner transformation $c_j=\left[\prod_{m=1}^{j-1}\sigma_m^z\right]\sigma_j^+$, which, up to an inconsequential constant energy shift, leads to the Hamiltonian
\begin{align}\nonumber
H =&\,- \sum_j \bigg[\frac{J+\Delta}{2}\sigma_j^x \sigma_{j+1}^x+\frac{J-\Delta}{2}\sigma_j^y \sigma_{j+1}^y\\\label{H_XYZ}
& +\frac{U}{4} \sigma_j^z\sigma_{j+1}^z+\bigg(h-\frac{U}{2}\bigg) \sigma_j^z \bigg],
\end{align}
This is the XYZ chain in a magnetic field along the $z$-direction. It generically possesses a $\mathrm{Z}_2$ symmetry due to invariance upon a $\pi$-rotation around the $z$-axis, which can be promoted to a $\mathrm{U}(1)$ symmetry for $\Delta=0$ (XXZ chain in a magnetic field where total $z$-magnetization is conserved), and even further to an $\mathrm{SU}(2)$ symmetry if additionally $U=2h=2J$ (Heisenberg chain). Another promotion occurs to $\mathrm{Z}_2\times\mathrm{Z}_2$ symmetry (due to invariance upon a $\pi$-rotation around each axis) when $U=2h$ for generic values of $J$ and $\Delta$. The XYZ model reduces to the paradigmatic nearest-neighbor transverse-field Ising chain (or, equivalently, the Kitaev chain at equal pairing and hopping strengths) for $\lvert J\rvert=\lvert\Delta\rvert$ and $U=0$, which also has a $\mathrm{Z}_2$ symmetry. The Hamiltonian~\eqref{H_XYZ} at zero magnetic-field strength ($U=2h$) and under periodic boundary conditions has been solved by relating it to the classical two-dimensional eight-vertex model\cite{Sutherland1970,Baxter1971} and through the algebraic Bethe Ansatz.\cite{Cao2014}

\bibliography{BIB_1}
\end{document}